%% file: main.tex
\documentclass[camera,letterpaper,nomarginnotes,nonarrowgutter]{jpaper}

\input{sections/packages}
\input{sections/commands}

\def\BibTeX{{\rm B\kern-.05em{\sc i\kern-.025em b}\kern-.08em
    T\kern-.1667em\lower.7ex\hbox{E}\kern-.125emX}}

\newcites{supp}{Supplementary References}

\begin{document}
\title{\ltitle}

\newcommand{\affilETH}[0]{\small {$^1$}}
\newcommand{\affilIntel}[0]{\small {$^2$}}
\newcommand{\affilCMU}[0]{\small {$^3$}}
\newcommand{\affilTUDelft}[0]{\small {$^4$}}
\author{
\vspace{-18pt}\\%
{Can Firtina\affilETH{}}\quad%
{Kamlesh Pillai\affilIntel{}}\quad%
{Gurpreet S. Kalsi\affilIntel{}}\quad%
{Bharathwaj Suresh\affilIntel{}}\quad%
{Damla Senol Cali\affilCMU{}}
\vspace{-1pt}\\%
\fontsize{11}{12}\selectfont%
{Jeremie S. Kim\affilETH{}}\quad%
{Taha Shahroodi\affilTUDelft{}}\quad%
{Meryem Banu Cavlak\affilETH{}}\quad%
{Joel Lindegger\affilETH{}}\quad%
{Mohammed Alser\affilETH{}}
\vspace{-1pt}\\%
\fontsize{11}{12}\selectfont%
{Juan Gómez Luna\affilETH{}}\quad%
{Sreenivas Subramoney\affilIntel{}}\quad%
{Onur Mutlu\affilETH{}}%
\vspace{-1pt}\\%
\vspace{-3pt}\\%
\affilETH\emph{ETH Zurich}%
\qquad\quad%
\affilIntel\emph{Intel Labs}%
\qquad\quad%
\affilCMU\emph{Carnegie Mellon University}%
\qquad\quad%
\affilTUDelft\emph{TU Delft}%
\vspace{-12pt}
}

\maketitle
\thispagestyle{plain}


\input{sections/0_abstract}
\input{sections/1_introduction.tex}
\input{sections/2_background}
\input{sections/3_motivation}
\input{sections/4_aphmm}

\input{sections/5_evaluation}
\input{sections/6_related-work}
\input{sections/7_conclusion}
\section*{Acknowledgments} 
We thank the SAFARI group members and Intel Labs for feedback and the stimulating intellectual environment. We acknowledge the generous gifts and support provided by our industrial partners: Intel, Google, Huawei, Microsoft, VMware, and the Semiconductor Research Corporation. This work is also partially supported by the European Union’s Horizon programme for research and innovation [101047160 - BioPIM] and the Swiss National Science Foundation (SNSF) [200021\_213084].

\bibliographystyle{IEEEtran}
\bibliography{main}

\setstretch{1}
\input{sections/supp}
\end{document}

%% file: sections/packages.tex
\usepackage{listings}
\usepackage{fancyhdr}

\usepackage{datetime}

\usepackage{cite}
\usepackage{amsmath,amssymb,amsfonts}
\usepackage{algorithmic}
\usepackage{graphicx}
\usepackage{textcomp}
\usepackage{xcolor}
\usepackage{tikz}

\usepackage[utf8]{inputenc}
\usepackage[T1]{fontenc}

\usepackage{booktabs} 
\usepackage{setspace}
\usepackage[italic]{mathastext}
\usepackage{array}
\usepackage{titlesec}
\usepackage[normalem]{ulem}
\usepackage{multirow}
\usepackage{multicol}
\usepackage{color}
\usepackage[font={small,bf}]{caption}
\usepackage{float}
\usepackage[font={small,bf}]{subcaption}
\usepackage[linesnumbered,ruled]{algorithm2e}
\usepackage{makecell}

\usepackage{pifont}
\usepackage[]{microtype}
            
\usepackage{todonotes} 
\usepackage{marginnote} 
\usepackage{pifont}
\let\marginpar\marginnote
\usepackage[bookmarks=true,breaklinks=true,colorlinks,linkcolor=black,citecolor=blue,urlcolor=black]{hyperref}
\usepackage[resetlabels]{multibib}

%% file: sections/commands.tex
\newcommand\proposal{ApHMM\xspace}
\newcommand\ltitle{ApHMM: Accelerating Profile Hidden Markov Models for Fast and Energy-Efficient Genome Analysis}

\newcommand\contb{Control Block\xspace}
\newcommand\compb{Compute Block\xspace}
\newcommand\pars{\texttt{Parameters}\xspace}
\newcommand\datacont{\texttt{Data Control}\xspace}
\newcommand\hif{\texttt{Histogram Filter}\xspace}
\newcommand\pe{\texttt{PE}\xspace}
\newcommand\pes{\texttt{PE}s\xspace}
\newcommand\peg{\texttt{PE} Group\xspace}
\newcommand\pegs{\texttt{PE} Groups\xspace}

\newcommand\luts{\texttt{LUT}s\xspace}

\newcommand\dotpt{\texttt{Dot Product Tree}\xspace}
\newcommand\accum{\texttt{Accumulator}\xspace}
\newcommand\redt{\texttt{Reduction Tree}\xspace}
\newcommand\temul{\texttt{TE MUL}\xspace}
\newcommand\ut{\texttt{UT}\xspace}
\newcommand\uts{\texttt{UTs}\xspace}
\newcommand\ue{\texttt{UE}\xspace}
\newcommand\ues{\texttt{UEs}\xspace}

\newcommand\pbaumcpua{$260.03\times$\xspace}

\newcommand\pbaumcpuc{$15.55\times$\xspace}
\newcommand\pbaumgpua{$5.34\times$\xspace}

\newcommand\pbaumgpud{$1.83\times$\xspace}
\newcommand\pbaumfpga{$27.97\times$\xspace}

\newcommand\pbaumgpucomp{$2.02\times$\xspace}

\newcommand\ebaumcpua{$2474.09\times$\xspace}
\newcommand\ebaumgpua{$2622.94\times$\xspace}

\newcommand\ebaumgpud{$896.70\times$\xspace}

\newcommand\perrcpua{$59.94\times$\xspace}

\newcommand\perrcpuc{$2.66\times$\xspace}
\newcommand\perrgpua{$2.09\times$\xspace}

\newcommand\perrgpud{$1.29\times$\xspace}
\newcommand\perrfpga{$7.21\times$\xspace}

\newcommand\eerrcpua{$64.24\times$\xspace}
\newcommand\eerrgpua{$115.46\times$\xspace}

\newcommand\eerrgpud{$71.28\times$\xspace}

\newcommand\pprocpua{$1.75\times$\xspace}

\newcommand\pprocpuc{$1.61\times$\xspace}
\newcommand\pprofpga{$1.03\times$\xspace}

\newcommand\eprocpua{$1.75\times$\xspace}

\newcommand\pmsacpua{$1.95\times$\xspace}
\newcommand\pmsafpga{$1.03\times$\xspace}

\newcommand\emsacpua{$1.96\times$\xspace}

\newcommand{\circlednumber}[1]{\raisebox{0.1pt}{
  \protect\tikz[baseline=(myanchor.base)]{
  \protect\node[circle,fill=.,inner sep=0.8pt] (myanchor) {\color{-.}\small \textbf{#1}};}%
}}

\newcommand{\circlednumberb}[1]{\raisebox{0.1pt}{
  \protect\tikz[baseline=(myanchor.base)]{
  \protect\node[circle,fill=.,inner sep=0.8pt] (myanchor) {\color{-.}\scriptsize \textbf{#1}};}%
}}

\definecolor{lightblue}{rgb}{0.980, 0.956, 0.623}
\definecolor{lightyellow}{rgb}{0.980, 0.956, 0.623}
\definecolor{darkpink}{rgb}{0.88, 0.28, 0.54}
\definecolor{forestgreen}{rgb}{0.0, 0.27, 0.13}
\definecolor{amber}{rgb}{1.0, 0.49, 0.0}

\newcommand{\head}[1]{{\noindent\textbf{#1.}\xspace}} 

\makeatletter
\g@addto@macro{\normalsize}{%
  \setlength{\abovedisplayskip}{2pt plus 1pt minus 1pt}
  \setlength{\belowdisplayskip}{2pt plus 1pt minus 1pt}
  \setlength{\intextsep}{2pt plus 1pt minus 1pt}
  \setlength{\textfloatsep}{3pt plus 1pt minus 1pt}
  \setlength{\dbltextfloatsep}{3pt plus 1pt minus 1pt}
  \setlength{\skip\footins}{4pt plus 1pt minus 1pt}
}
\def\BibTeX{{\rm B\kern-.05em{\sc i\kern-.025em b}\kern-.08em
    T\kern-.1667em\lower.7ex\hbox{E}\kern-.125emX}}

\def\UrlBreaks{\do\/\do-\/\do.\/\do:}

\expandafter\def\expandafter\UrlBreaks\expandafter{\UrlBreaks
  \do\a\do\b\do\c\do\d\do\e\do\f\do\g\do\h\do\i\do\j
  \do\k\do\l\do\m\do\n\do\o\do\p\do\q\do\r\do\s\do\t
  \do\u\do\v\do\w\do\x\do\y\do\z\do\A\do\B\do\C\do\D
  \do\E\do\F\do\G\do\H\do\I\do\J\do\K\do\L\do\M\do\N
  \do\O\do\P\do\Q\do\R\do\S\do\T\do\U\do\V\do\W\do\X
  \do\Y\do\Z}
  
\newcommand{\squishlist}{
 \begin{list}{$\circ$}
  { \setlength{\itemsep}{0pt}
     \setlength{\parsep}{0pt}
     \setlength{\topsep}{0pt}
     \setlength{\partopsep}{0pt}
     \setlength{\leftmargin}{1em}
     \setlength{\labelwidth}{1em}
     \setlength{\labelsep}{0.5em} } }

\newcommand{\squishsublist}{
\begin{list}{$\rightarrow$}
 { \setlength{\itemsep}{0pt}
    \setlength{\parsep}{0pt}
    \setlength{\topsep}{-10em}
    \setlength{\partopsep}{-3pt}
    \setlength{\leftmargin}{1em}
    \setlength{\labelwidth}{1em}
    \setlength{\labelsep}{0.5em} } }

\newcommand{\squishend}{
  \end{list}  }

\usepackage{titlesec}
\titlespacing*{\section}{0pt}{0.3ex}{0.1ex} 
\titlespacing*{\subsection}{0pt}{0.3ex}{0.3ex} 
\titlespacing*{\subsubsection}{0pt}{0.3ex}{0.3ex} 

%% file: sections/0_abstract.tex
\begin{abstract}
Profile hidden Markov models (pHMMs) are widely employed in various bioinformatics applications to identify similarities between biological sequences, such as DNA or protein sequences. In pHMMs, sequences are represented as graph structures, where states and edges capture modifications (i.e., insertions, deletions, and substitutions) by assigning probabilities to them. These probabilities are subsequently used to compute the similarity score between a sequence and a pHMM graph. The Baum-Welch algorithm, a prevalent and highly accurate method, utilizes these probabilities to optimize and compute similarity scores. Accurate computation of these probabilities is essential for the correct identification of sequence similarities. However, the Baum-Welch algorithm is computationally intensive, and existing solutions offer either software-only or hardware-only approaches with fixed pHMM designs. When we analyze state-of-the-art works, we identify an urgent need for a flexible, high-performance, and energy-efficient hardware-software co-design to address the major inefficiencies in the Baum-Welch algorithm for pHMMs.

We introduce \emph{\proposal}, the \emph{first} flexible acceleration framework designed to significantly reduce both computational and energy overheads associated with the Baum-Welch algorithm for pHMMs. \proposal employs hardware-software co-design to tackle the major inefficiencies in the Baum-Welch algorithm by 1)~designing flexible hardware to accommodate various pHMM designs, 2)~exploiting predictable data dependency patterns through on-chip memory with memoization techniques, 3)~rapidly filtering out negligible computations using a hardware-based filter, and 4)~minimizing redundant computations.

\proposal achieves substantial speedups of \pbaumcpuc~-~\pbaumcpua, \pbaumgpud~-~\pbaumgpua, and \pbaumfpga when compared to CPU, GPU, and FPGA implementations of the Baum-Welch algorithm, respectively. \proposal outperforms state-of-the-art CPU implementations in three key bioinformatics applications: 1)~error correction, 2)~protein family search, and 3)~multiple sequence alignment, by \perrgpud~-~\perrcpua, \pprofpga~-~\pprocpua, and \pmsafpga~-~\pmsacpua, respectively, while improving their energy efficiency by \eerrcpua~-~\eerrgpua, \eprocpua, \emsacpua.
\end{abstract}

%% file: sections/1_introduction.tex
\section{Introduction} \label{sec:introduction}
Hidden Markov Models (HMMs) are useful for calculating the probability of a sequence of previously unknown (hidden) events (e.g., the weather condition) given observed events (e.g., clothing choice of a person)~\cite{eddy_what_2004}. To calculate the probability, HMMs use a graph structure where a sequence of nodes (i.e., states) are visited based on the series of observations with a certain probability associated with visiting a state from another. HMMs are very efficient in decoding the continuous and discrete series of events in many applications~\cite{mor_systematic_2021} such as speech recognition~\cite{mor_systematic_2021, mustafa_comparative_2019, mao_revisiting_2019, hamidi_interactive_2018, xue_novel_2018, li_hybrid_2013, patel_speech_2010}, text classification~\cite{nasim_sentiment_2020, kang_opinion_2018, zeinali_text-dependent_2017, ahmad_open-vocabulary_2016, vieira_t-hmm_2014}, gesture recognition~\cite{moreira_acoustic_2020, sinha_computer_2019, haid_inertial-based_2019, calin_gesture_2016, deo_-vehicle_2016, malysa_hidden_2016, nguyen-duc-thanh_two-stage_2012, shrivastava_hidden_2013}, and bioinformatics~\cite{liang_bayesian_2007, boufounos_basecalling_2004, narasimhan_bcftoolsroh_2016, yin_args-oap_2018, tamposis_semi-supervised_2019, zhang_fish_2014, eddy_profile_1998, huang_hardware_2017, wu_173_2020}. The graph structures (i.e., designs) of HMMs are typically tailored for each application, which defines the roles and probabilities of the states and edges connecting these states, called \emph{transitions}. One important special design of HMMs is known as the \emph{profile Hidden Markov Model} (pHMM) design~\cite{eddy_profile_1998}, which is commonly adopted in bioinformatics~\cite{baldi_hidden_1994, bateman_pfam_2002, zhang_profile_2003, sgourakis_method_2005, friedrich_modelling_2006, steinegger_hh-suite3_2019, durbin_biological_1998, edgar_coach_2004, madera_profile_2008, eddy_accelerated_2011, wheeler_dfam_2012, firtina_hercules_2018, firtina_apollo_2020, lanyue_long_2020, yoon_hidden_2009}, malware detection~\cite{ali_profile_2022, sasidharan_prodroid_2021, liu_adversarial_2019, pranamulia_profile_2017, ravi_behavior-based_2013, attaluri_profile_2009} and pattern matching~\cite{riddell_reliable_2022, kazantzidis_profile_2018, saadi_framework_2016, ding_skeleton-based_2015, liu_characterizing_2015, liu_who_2009}.

\begin{figure}[ht]
\centering
\includegraphics[width=\columnwidth]{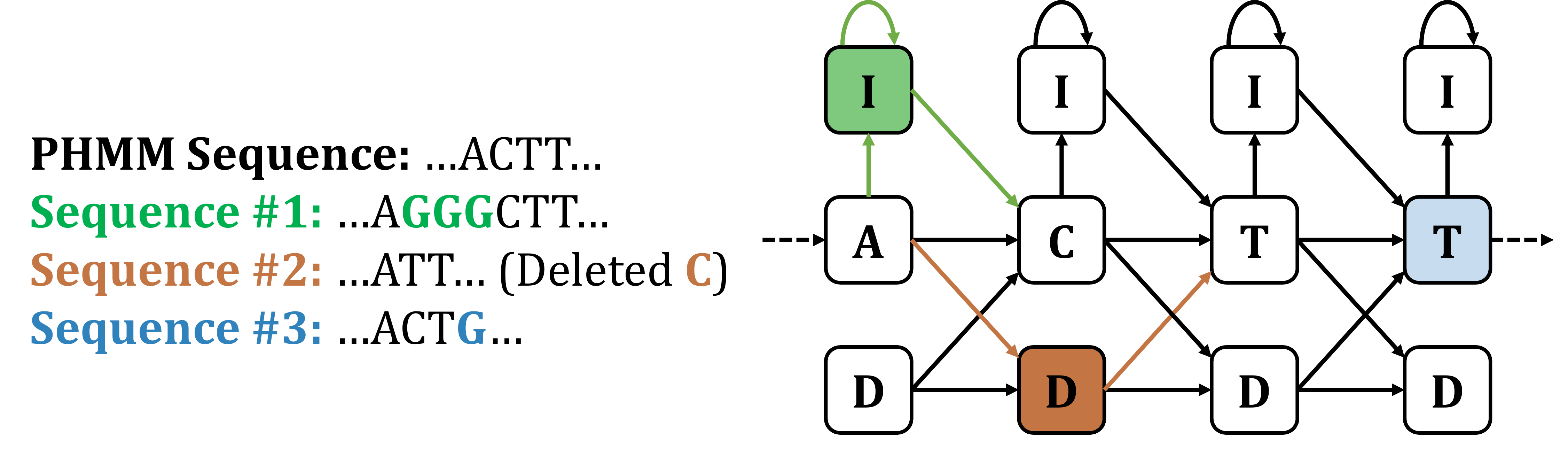}
\caption{Portion of an example pHMM design that represents a DNA sequence (\texttt{PHMM Sequence}).
Differences between \texttt{PHMM Sequence} and \texttt{Sequences \#1}, \texttt{\#2}, and \texttt{\#3} are highlighted by colors. Highlighted transitions and states identify each corresponding difference.}
\label{fig:phmm}
\end{figure}

Identifying differences between biological sequences (e.g., DNA sequences) is an essential step in bioinformatics applications to understand the effects of these differences (e.g., genetic variations and their relations to certain diseases)~\cite{alser_technology_2021, firtina_rawhash_2023, firtina_rawhash2_2023, lindegger_rawalign_2023, alser_going_2022, alser_accelerating_2020, singh_fpga-based_2021, alser_sneakysnake_2020, angizi_pim-aligner_2020, goenka_segalign_2020, senol_cali_genasm_2020, turakhia_darwin_2018, kim_grim-filter_2018, mansouri_ghiasi_genstore_2022, nag_gencache_2019, firtina_blend_2021, cali_segram_2022, kim_airlift_2021, kim_fastremap_2022}. PHMMs enable efficient and accurate identification of differences by comparing sequences to a few graphs that represent a group of sequences rather than comparing many sequences to each other, which is computationally very costly and requires special hardware and software optimizations~\cite{alser_technology_2021, cali_segram_2022}. Figure~\ref{fig:phmm} illustrates a \emph{traditional} design of pHMMs. A pHMM represents a single or many sequences with a graph structure using states and transitions. There are three types of states for each character of a sequence that a pHMM graph represents: insertion (\texttt{I}), match or mismatch (\texttt{M}), and deletion (\texttt{D}) states. Each state accounts for a certain difference or a match between a graph and an input sequence at a particular position. For example, the \texttt{I} states recognize insertions in an input sequence missing from the pHMM graph at a position. Many bioinformatics applications use pHMM graphs rather than directly comparing sequences to avoid the high cost of many sequence comparisons. The applications that use pHMMs include the protein family search~\cite{baldi_hidden_1994, bateman_pfam_2002, zhang_profile_2003, friedrich_modelling_2006, steinegger_hh-suite3_2019, soding_hhpred_2005, finn_pfam_2010, madera_comparison_2002}, the multiple sequence alignment (MSA)~\cite{durbin_biological_1998, edgar_coach_2004, madera_profile_2008, attaluri_profile_2009, eddy_accelerated_2011, wheeler_dfam_2012, sgourakis_method_2005, steinegger_hh-suite3_2019, mulia_profile_2012, pei_promals_2007, edgar_satchmo_2003}, and error correction~\cite{firtina_hercules_2018, firtina_apollo_2020, lanyue_long_2020}.

To accurately model and compare DNA or protein sequences using pHMMs, assigning accurate probabilities to states and transitions is essential. PHMMs allow updating these probabilities to accurately fit the observed biological sequences to the pHMM graph. Probabilities are adjusted during the \emph{training} step. The training step aims to maximize the probability of observing the input biological sequences in a given pHMM, also known as \emph{likelihood maximization}. There are several algorithms that perform such a maximization in pHMMs~\cite{baum_inequality_1972, scott_bayesian_2002, lewis_bayesian_2008, rezaei_generalized_2013}. The Baum-Welch algorithm~\cite{baum_inequality_1972} is commonly used to calculate the likelihood maximization~\cite{boussemart_comparing_2009} as it is highly accurate and scalable to real-size problems (e.g., large protein families)~\cite{lewis_bayesian_2008}. The next step is \emph{inference}, which aims to identify either 1)~the similarity of an input observation sequence to a pHMM graph or 2)~the sequence with the highest similarity to the pHMM graph, which is known as the \emph{consensus sequence} of the pHMM graph and used for error correction in biological sequences. Parts of the Baum-Welch algorithm can be used for calculating the similarity of an input sequence in the inference step.

Despite its advantages, the Baum-Welch algorithm is a computationally expensive method~\cite{lyngso_consensus_2002, kahsay_quasi-consensus-based_2005} due to the nature of its dynamic programming approach. Several works~\cite{eddy_accelerated_2011, ren_fpga_2015, pietras_fpga_2017, yu_gpu-accelerated_2014, soiman_parallel_2014} aim to accelerate either the entire or smaller parts of the Baum-Welch algorithm for HMMs or pHMMs to mitigate the high computational costs. While these works can improve the performance for executing the Baum-Welch algorithm, they either 1)~provide software- or hardware-only solutions for a fixed pHMM design or 2)~are completely oblivious to the pHMM design.

To identify the inefficiencies in using pHMMs with the Baum-Welch algorithm, we analyze the state-of-the-art implementations of three pHMM-based bioinformatics applications: 1)~error correction, 2)~protein family search, and 3)~multiple sequence alignment (Section~\ref{sec:motivation}). We make six key observations. 1)~The Baum-Welch algorithm causes a significant computational overhead in the pHMM applications as it constitutes at least around $50\%$ of the total execution time of these applications. 2)~SIMD-based approaches cannot fully vectorize the floating-point operations. 3)~A significant portion of floating-point operations is redundant in the training step due to a lack of a mechanism for reusing the same products. 4)~Existing strategies for filtering out the negligible states from the computation are costly despite their advantages. 5)~The spatial locality inherent in pHMMs cannot be exploited in generic HMM-based accelerators and applications as these accelerators and applications are oblivious to the design of HMMs. 6)~The Baum-Welch algorithm is the main source of computational overhead even for the non-genomic application we evaluate. Unfortunately, software- or hardware-only solutions cannot solve these inefficiencies. These observations show a pressing need for a flexible, high-performant, and energy-efficient hardware-software co-design to efficiently and effectively solve these inefficiencies in the Baum-Welch algorithm for pHMMs.

Our \textbf{goal} is to accelerate the Baum-Welch algorithm while eliminating the inefficiencies when executing the Baum-Welch algorithm for pHMMs. To this end, we propose \proposal, the \emph{first} flexible hardware-software co-designed acceleration framework that can significantly reduce the computational and energy overheads of the Baum-Welch algorithm for pHMMs. \proposal is built on four \textbf{key mechanisms}. First, \proposal is highly flexible and can use different pHMM designs to change certain parameter choices to enable the adoption of \proposal for many pHMM-based applications. This enables 1)~additional support for pHMM-based error correction~\cite{lanyue_long_2020, firtina_hercules_2018, firtina_apollo_2020} that traditional pHMM design cannot efficiently and accurately support~\cite{firtina_hercules_2018}. Second, \proposal exploits the spatial locality that pHMMs provide with the Baum-Welch algorithm by efficiently utilizing on-chip memories with memoizing techniques. Third, \proposal efficiently eliminates negligible computations with a hardware-based filter design. Fourth, \proposal avoids redundant floating-point operations by 1)~providing a mechanism for efficiently reusing the most common products of multiplications in lookup tables (LUTs) and 2)~identifying pipelining and broadcasting opportunities where certain computations are moved between multiple steps in the Baum-Welch algorithm without extra storage or computational overheads. The fourth mechanism includes our software optimizations, while on-chip memory and a hardware-based filter require a special and efficient hardware design.

To evaluate \proposal, we 1)~design a flexible hardware-software co-designed acceleration framework in an accelerator and 2)~implement the software optimizations for GPUs. We evaluate the performance and energy efficiency of \proposal for executing 1)~the Baum-Welch algorithm and 2)~several pHMM-based applications and compare \proposal to the corresponding CPU, GPU, and FPGA baselines. First, our extensive evaluations show that \proposal provides significant 1)~speedup for executing the Baum-Welch algorithm by \pbaumcpuc - \pbaumcpua (CPU), \pbaumgpud - \pbaumgpua (GPU), and \pbaumfpga (FPGA) and 2)~energy efficiency by \ebaumcpua (CPU) and \ebaumgpud-\ebaumgpua (GPU). Second, \proposal improves the overall runtime of the pHMM-based applications, error correction, protein family search, and MSA by \perrgpud - \perrcpua, \pprofpga - \pprocpua, and \pmsafpga - \pmsacpua and reduces their overall energy consumption by \eerrcpua - \eerrgpua, \eprocpua, \emsacpua over their state-of-the-art CPU, GPU, and FPGA implementations, respectively. We make the following \textbf{key contributions}:

\vspace{-0.1cm}
\begin{itemize}
    \item We introduce \proposal, the first flexible hardware-software co-designed framework to accelerate pHMMs. We show that our framework can be used at least for three bioinformatics applications: 1)~error correction, 2)~protein family search, and 3)~multiple sequence alignment.
    \item We provide \proposal-GPU, the first GPU implementation of the Baum-Welch algorithm for pHMMs, which includes our software optimizations. 
    \item We identify key inefficiencies in the state-of-the-art pHMM applications and provide key mechanisms with efficient hardware and software optimizations to significantly reduce the computational and energy overhead of the Baum-Welch algorithm for pHMMs.
    \item We show that \proposal provides significant speedups and energy reductions for executing the Baum-Welch algorithm compared to the CPU, GPU, and FPGA implementations, while \proposal-GPU performs better than the state-of-the-art GPU implementation.
    \item We provide the source code of our software optimizations, \proposal-GPU, as implemented in an error correction application. The source code is available at \url{https://github.com/CMU-SAFARI/ApHMM-GPU}.
 \end{itemize}

%% file: sections/2_background.tex
\section{Background} \label{sec:background}
\subsection{Profile Hidden Markov Models (pHMMs)}\label{subsec:phmms}
\head{High-level Overview}
Figure~\ref{fig:phmm} shows the traditional structure of pHMMs. PHMMs represent a sequence or a group of sequences using a certain graph structure with a fixed number of nodes for every character of represented sequences. Visiting nodes, called \emph{states}, via directed edges, called \emph{transitions}, are associated with probabilities to identify differences at any position between the represented sequences and other sequences. States emit one of the characters from the defined alphabet of the biological sequence (e.g., A, C, T, and G in DNA sequences) with a certain probability. Transitions preserve the correct order of the represented sequences and allow making modifications to these sequences. We explain the detailed structure of pHMMs in Supplemental Section 1.

\subsection{The Baum-Welch Algorithm}\label{subsec:BW_alg}
The probabilities associated with transitions and states are essential for identifying similarities between sequences. The Baum-Welch algorithm provides a set of equations to update and use these probabilities accurately. To calculate the similarity score of input observation sequences in a pHMM graph, the Baum-Welch algorithm~\cite{baum_inequality_1972} solves an expectation-maximization problem~\cite{moon_expectation-maximization_1996, tavanaei_training_2018, lindberg_petro-elastic_2015, hubin_adaptive_2019}, where the expectation step calculates the statistical values based on an input sequence to train the probabilities of pHMMs. To this end, the algorithm performs the expectation-maximization based on an observation sequence $S$ for the pHMM graph $G(V,A)$ in three steps: 1)~forward calculation, 2)~backward calculation, and 3)~parameter updates.

\head{Forward Calculation}
The goal of the forward calculation is to compute the probability of observing sequence $S$ when we compare it with the sequence $S_{G}$ that the pHMM graph $G(V,A)$ represents. Equation~\ref{eq:forward} shows the calculation of the forward value $F_{t}(i)$ of state $v_i$ for character $S[t]$. The forward value, $F_{t}(i)$, represents the likelihood of emitting the character $S[t]$ at position $t$ of $S$ in state $v_{i}$ given that \emph{all} previous characters $S[1 \dots t-1]$ are emitted by following an unknown path \emph{forward} that leads to state $v_{i}$. $F_{t}(i)$ is calculated for all states $v_i \in V$ and for all characters of $S$. Although $t$ represents the position of the character of $S$, we use the \emph{timestamp} term for $t$ for the remainder of this paper. To represent transition and emission probabilities, we use the $\alpha_{ji}$ and $e_{S[t]}(v_{i})$ notations as we define in Supplemental Section 1.2.

\begin{equation}
F_{t}(i) = \sum_{j \in V} F_{t-1}(j) \alpha_{ji} e_{S[t]}(v_{i}) \enspace i \in V, \enspace 1 < t \leq n_{S} \tag{1} \label{eq:forward}
\end{equation}

\head{Backward Calculation}
The goal of the backward calculation is to compute the probability of observing sequence $S$ when we compare $S$ and $S_G$ from their last characters to the first characters. Equation~\ref{eq:backward} shows the calculation of the backward value $B_{t}(i)$ of state $v_i$ for character $S[t]$. The backward value, $B_{t}(i)$, represents the likelihood of emitting $S[t]$ in state $v_{i}$ given that \emph{all} further characters $S[t+1 \dots n_{S}]$ are emitted by following an unknown path \emph{backwards} (i.e., taking transitions in reverse order). $B_{t}(i)$ is calculated for all states $v_i \in V$ and for all characters of $S$.

\begin{equation}
B_{t}(i) = \sum_{j \in V} B_{t+1}(j) \alpha_{ij} e_{S[t+1]}(v_{j}) \enspace i \in V, ~1 \leq t < n_{S} \tag{2} \label{eq:backward}
\end{equation}

\head{Parameter Updates}
The Baum-Welch algorithm uses the values that the forward and backward calculations generate for the observation sequence $S$ to update the emission and transition probabilities in $G(V, A)$. The parameter update procedure maximizes the similarity score of $S$ in $G(V, A)$. This procedure updates the parameters shown in Equations~\ref{eq:transition} and ~\ref{eq:emission}. The special $[S[t] = X]$ notation in Equation~\ref{eq:emission} is a conditional variable such that the variable returns $1$ if the character $X$ matches with the character $S[t]$ and returns $0$ otherwise.

\begin{equation}
\alpha^{*}_{ij} = \dfrac{\sum\limits_{t=1}^{n_{S}-1} \alpha_{ij}e_{S[t+1]}(v_{j})F_{t}(i)B_{t+1}(j)}
     {\sum\limits_{t=1}^{n_{S}-1}\sum\limits_{x \in V} \alpha_{ix}e_{S[t+1]}(v_{x})F_{t}(i)B_{t+1}(x)} \quad \forall \alpha_{ij} \in A
\tag{3} \label{eq:transition}
\end{equation}

\begin{equation}
e^{*}_{X}(v_{i}) = \dfrac{\sum\limits_{t=1}^{n_{S}} F_{t}(i)B_t(i)[S[t] = X]}
             {\sum\limits_{t=1}^{n_{S}} F_{t}(i)B_t(i)} \quad \forall X \in \Sigma, \forall i \in V \tag{4} \label{eq:emission}
\end{equation}

\subsection{Use Cases of Profile HMMs}
\head{Error Correction}
The goal of error correction is to locate the erroneous parts in DNA or genome sequences to replace these parts with more reliable sequences~\cite{vaser_fast_2017, hu_nextpolish_2020, huang_neuralpolish_2021, walker_pilon_2014, zimin_genome_2020, chin_nonhybrid_2013} to enable more accurate genome analysis (e.g., read mapping and genome assembly). Apollo~\cite{firtina_apollo_2020} is a recent error correction algorithm that takes an assembly sequence and a set of reads as input to correct the errors in an assembly. Apollo constructs a pHMM graph for an assembly sequence to correct the errors in two steps: 1)~training and 2)~inference. First, to correct erroneous parts in an assembly, Apollo uses reads as observations to train the pHMM graph with the Baum-Welch algorithm. Second, Apollo uses the Viterbi algorithm~\cite{viterbi_error_1967} to identify the consensus sequence from the trained pHMM, which translates into the corrected assembly sequence. Apollo uses a slightly modified design of pHMMs to avoid certain limitations associated with traditional pHMMs when generating the consensus sequences\cite{lyngso_consensus_2002, kahsay_quasi-consensus-based_2005}. The modified design avoids loops in the insertion states and uses transitions to account for deletions instead of deletion states. These modifications allow the pHMM-based error correction applications~\cite{lanyue_long_2020, firtina_hercules_2018, firtina_apollo_2020} to construct more accurate consensus sequences from pHMMs.

\head{Protein Family Search}
Classifying protein sequences into families is widely used to analyze the potential functions of the proteins of interest~\cite{mulder_tools_2001, jeffryes_rapid_2018, seo_deepfam_2018, vicedomini_multiple_2022, turjanski_natural_2018, bileschi_using_2022}. The protein family search finds the family of the protein sequence in existing protein databases. A pHMM usually represents one protein family in the database to avoid searching for many individual sequences. The protein sequence can then be assigned to a protein family based on the similarity score of the protein when compared to a pHMM in a database. This approach is used to search protein sequences in the Pfam database~\cite{mistry_pfam_2021}, where the HMMER~\cite{eddy_accelerated_2011} software suite is used to build HMMs and assign query sequences to the best fitting Pfam family.
Similar to the Pfam database, HMMER's protein family search tool is integrated into the European Bioinformatics Institute (EBI) website as a web tool.
The same approach is also used in several other important applications, such as classifying many genomic sequences into potential viral families~\cite{skewes-cox_profile_2014}.

\head{Multiple Sequence Alignment}
Multiple sequence alignment (MSA) detects the differences between several biological sequences. Dynamic programming algorithms can optimally find differences between genomic sequences, but the complexity of these algorithms increases drastically with the number of sequences~\cite{just_computational_2001, wang_complexity_1994}. To mitigate these computational problems, heuristics algorithms are used to obtain an approximate yet computationally efficient solution for multiple alignments of genomic sequences. PHMM-based approaches provide an efficient solution for MSA~\cite{chowdhury_review_2017}. The pHMM approaches, such as \emph{hmmalign}~\cite{eddy_accelerated_2011}, assign likelihoods to all possible combinations of differences between sequences to calculate the pairwise similarity scores using forward and backward calculations or other optimization methods (e.g., particle swarm optimization~\cite{zhan_probpfp_2019}). PHMM-based MSA approaches are mainly useful to avoid making redundant comparisons as a sequence can be compared to a pHMM graph, similar to the protein family search.

%% file: sections/3_motivation.tex
\section{Motivation and Goal} \label{sec:motivation}
\subsection{Sources of Inefficiencies}\label{subsec:inefficiencies}
To identify and understand the performance overheads of state-of-the-art pHMM-based applications, we thoroughly analyze existing tools for the three use cases of pHMM: 1)~Apollo~\cite{firtina_apollo_2020} for error correction, 2)~hmmsearch~\cite{eddy_accelerated_2011} for the protein family search, and 3)~hmmalign~\cite{eddy_accelerated_2011} for the multiple sequence alignment (MSA). We make six key observations based on our profiling with Intel VTune~\cite{profiler2022intel} and gprof~\cite{graham_gprof_2004}.

\begin{figure}[ht]
\centering
\includegraphics[width=\columnwidth]{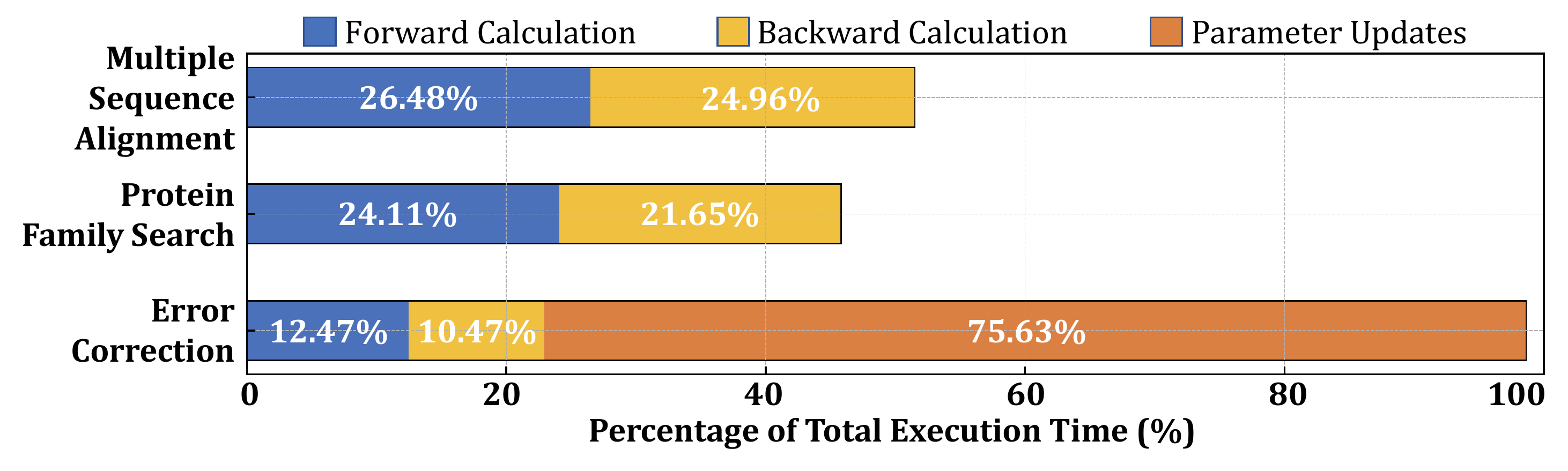}
\caption{Percentage of the total execution time for the three steps of the Baum-Welch algorithm}
\label{fig:motivation}
\end{figure}

\head{Observation 1: The Baum-Welch Algorithm causes a significant computational overhead} Figure~\ref{fig:motivation} shows the percentage of the execution time of all three steps in the Baum-Welch algorithm for the three bioinformatics applications. We find that the Baum-Welch algorithm causes a significant performance overhead for all three applications as the algorithm constitutes from 45.76\% up to $98.57\%$ of the total CPU execution time. Our profiling shows that these applications are mainly compute-bound. Forward and Backward calculations are the common steps in all three applications, whereas the Parameter Updates step is executed only for error correction. This is because the protein family search and MSA use the Forward and Backward calculations mainly for scoring between a sequence and a pHMM graph as a part of inference. We do \emph{not} include the cost of training for these applications as it is executed once or only a few times, such that the cost of training becomes insignificant compared to the frequently executed inference. However, the nature of error correction requires frequently performing both training and inference for every input sequence such that the cost of training is not negligible for this application. As a result, accelerating the entire Baum-Welch algorithm is key for accelerating the end-to-end performance of the applications.

\head{Observation 2: SIMD-based tools on CPU and GPUs provide suboptimal vectorization}
The Baum-Welch algorithm involves frequent floating-point multiplications and additions. To resolve performance issues, several CPU-based tools (e.g., hmmalign) use SIMD instructions. However, these tools exhibit poor SIMD utilization due to inadequate port utilization and low vector capacity usage (below 50\%). This suggests that CPU-based optimizations for floating-point operations, such as SIMD instructions, provide limited computational benefits for the Baum-Welch algorithm. We further investigate if the SIMD utilization in GPUs exhibits similar low utilization. To observe this, we profile our GPU work, \proposal-GPU, to execute the two main kernels in the application: Forward and Backward calculations. We observe that the Forward calculation suffers from low SIMD utilization (i.e., percentage of active threads per warp) of around 50\%, while the SIMD utilization of Backward calculations is usually close to 100\%. The nature of the GPU implementation iterates over all the states that have a connection to the state that the thread is working on. However, the number of states to iterate can substantially be different per thread during the Forward calculation as insertion and match states largely have a different number of \emph{incoming states}, which is not the case in Backward calculation. This imbalance causes high warp divergence during Forward calculation, reducing the SIMD utilization overall.

\head{Observation 3: A significant portion of floating-point operations is redundant} We observe that the same multiplications are repeatedly executed in the training step because certain floating-point values associated with transition and emission probabilities are mainly constant during training in error correction. Our profiling analysis with VTune shows that these redundant computations constitute around $22.7\%$ of the overall execution time when using the Baum-Welch algorithm for training in error correction.

\begin{figure}[ht]
\centering
\includegraphics[width=\columnwidth]{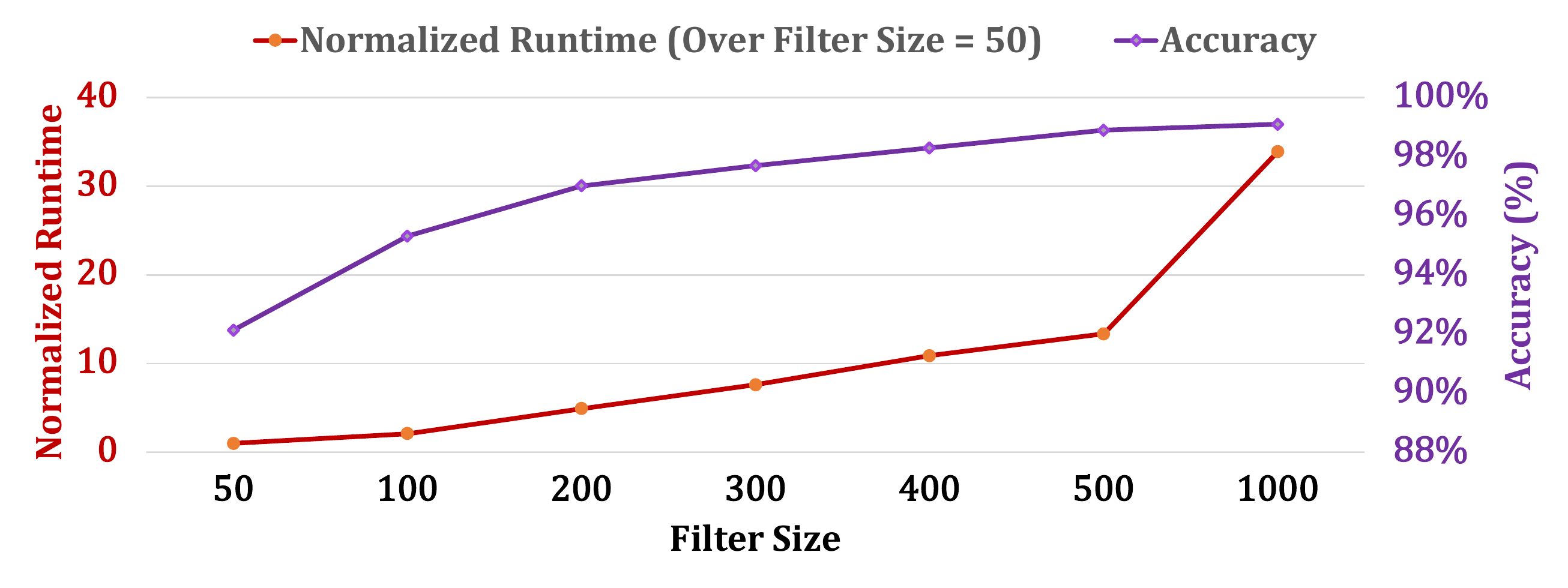}
\caption{Effect of the filter size on the runtime and the accuracy of the Baum-Welch algorithm}
\label{fig:motivation-filtersize}
\end{figure}

\head{Observation 4: Filtering the states is costly despite its advantages} The Baum-Welch algorithm requires performing many operations for a large number of states. These operations are repeated in many iterations, and the number of states can grow with each iteration. There are several approaches to keep the state space (i.e., number of states) near-constant to improve the performance or the space efficiency of the Baum-Welch algorithm~\cite{kirkpatrick_optimal_2012, miklos_linear_2005, grice_reduced_1997, wheeler_optimizing_2000, tarnas_reduced_1998, firtina_hercules_2018, firtina_apollo_2020}. A simple approach is to pick the \texttt{best-n} states that provide the highest scores at each iteration while the rest of the states are ignored in the next iteration, known as filtering~\cite{firtina_hercules_2018}. Figure~\ref{fig:motivation-filtersize} shows the relation between the filter size (i.e., the number of states picked as \texttt{best-n} states), runtime, and accuracy. Although the filtering approach is useful for reducing the runtime without significantly degrading the overall accuracy of the Baum-Welch algorithm, such an approach requires extra computations (e.g., sorting) to pick the \texttt{best-n} states. We find that such a filtering approach incurs non-negligible performance costs by constituting around $8.5\%$ of the overall execution time in the training step.

\begin{figure}[ht]
\centering
\includegraphics[width=\columnwidth]{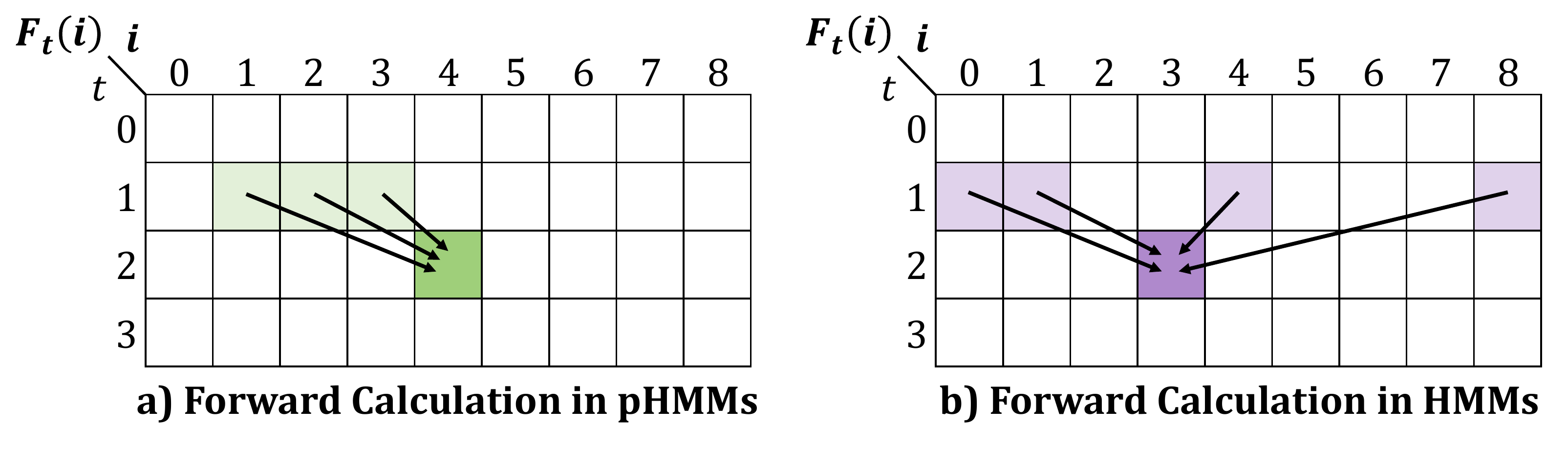}
\caption{Data dependency in pHMMs and HMMs}
\label{fig:motivation-datadependency}
\end{figure}

\head{Observation 5: HMM accelerators are suboptimal for accelerating pHMMs}
Generic HMMs do not require constraints on the connection between states (i.e., transitions) and the number of states. PHMMs are a special case for HMMs where transitions are predefined, and the number of states is determined based on the sequence that a pHMM graph represents. These design choices in HMMs and pHMMs affect the data dependency pattern when executing the Baum-Welch Algorithm. Figure~\ref{fig:motivation-datadependency} shows an example of the data dependency patterns in pHMMs and HMMs when executing the Baum-Welch algorithm. We observe that although HMMs and pHMMs provide similar temporal localities (e.g., only the values from the previous iteration are used), pHMMs provide better spatial localities with their constrained design. This observation suggests that HMM-based accelerators cannot fully exploit the spatial localities that pHMMs provide as they are oblivious to the design of pHMMs.

\head{Observation 6: Non-genomics pHMM-based applications suffer from the computational overhead of the Baum-Welch algorithm} Among many non-genomics pHMM-based implementations~\cite{ali_profile_2022, sasidharan_prodroid_2021, liu_adversarial_2019, pranamulia_profile_2017, ravi_behavior-based_2013, attaluri_profile_2009, riddell_reliable_2022, kazantzidis_profile_2018, saadi_framework_2016, ding_skeleton-based_2015, liu_characterizing_2015, liu_who_2009}, we analyze the available CPU implementation of a recent pattern-matching application that uses pHMMs~\cite{riddell_reliable_2022}. Our initial analysis shows that almost the entire execution time ($98\%$) of this application is spent on the Forward calculation, and it takes significantly longer to execute a relatively small dataset compared to the bioinformatics applications.

Many applications use either the entire or parts of the Baum-Welch algorithm for training the probabilities of HMMs and pHMMs~\cite{bateman_pfam_2002, zhang_profile_2003, sgourakis_method_2005, friedrich_modelling_2006, steinegger_hh-suite3_2019, attaluri_profile_2009, eddy_accelerated_2011, firtina_hercules_2018, firtina_apollo_2020, lanyue_long_2020, boufounos_basecalling_2004, chen_detecting_2016, ali_profile_2022, sasidharan_prodroid_2021, ravi_behavior-based_2013, riddell_reliable_2022, kazantzidis_profile_2018, ding_skeleton-based_2015}.
However, due to computational inefficiencies, the Baum-Welch algorithm can result in significant performance overheads on these applications.
Solving the inefficiencies in the Baum-Welch algorithm is mainly important for services that frequently use these applications, such as the EBI website using HMMER for searching protein sequences in protein databases~\cite{madeira_embl-ebi_2019}. Based on the latest report in 2018, there have been more than 28 million HMMER queries on the EBI website within two years (2016-2017)~\cite{potter_hmmer_2018}. These queries execute parts of the Baum-Welch algorithm more than 38,000 times daily. Such frequent usage leads to significant waste in compute cycles and energy due to the inefficiencies in the Baum-Welch algorithm.

While the Baum-Welch algorithm is computationally intensive and can consume a significant portion of the runtime and energy in various applications, these applications are often run multiple times as part of routine analyses or when new data becomes available. For error correction, the assembly of a particular genome can be reconstructed and corrected multiple times if additional sequencing data for the genome becomes available or if the tools used in the assembly construction pipeline are updated or replaced. For the protein family search and the multiple sequence alignment, protein sequencing data is frequently used multiple times due to regular updates in databases like the Pfam database~\cite{el-gebali_pfam_2019, mistry_pfam_2021}. These updates can generate new insights~\cite{li_refseq_2021}, such as more accurate reannotation of genes in assemblies~\cite{lorenzi_new_2010}. This frequent use of sequenced data can make using the Baum-Welch algorithm a time and energy-consuming process in the overall sequencing data analysis pipeline. Improving the efficiency of the Baum-Welch algorithm can significantly reduce both the compute cycles and energy consumption, especially in use cases where sequencing data is used multiple times.

\subsection{Goal}

Based on our observations, we find that we need to have a specialized, flexible, high-performant, and energy-efficient design to \circlednumber{$1$} support different pHMM designs with specialized compute units for each step in the Baum-Welch algorithm, \circlednumber{$2$} eliminate redundant operations by enabling efficient reuse of the common multiplication products, \circlednumber{$3$} exploit spatiotemporal locality in on-chip memory, and \circlednumber{$4$} perform efficient filtering. Such a design has the potential to significantly reduce the computational and energy overhead of the applications that use the Baum-Welch algorithm in pHMMs. Unfortunately, software- or hardware-only solutions cannot solve these inefficiencies easily. There is a pressing need to develop a hardware-software co-designed and flexible acceleration framework for several pHMM-based applications that use the Baum-Welch algorithm.

In this work, our \textbf{goal} is to reduce computational and energy overhead of the pHMMs-based applications that use the Baum-Welch algorithm with a flexible, high-performance, energy-efficient hardware-software co-designed acceleration framework. To this end, we propose \proposal, the \emph{first} highly flexible, high-performant, and energy-efficient accelerator that can support different pHMM designs to accelerate wide-range pHMM-based applications.

%% file: sections/4_aphmm.tex
\section{ApHMM Design} \label{sec:aphmm}
\subsection{Microarchitecture Overview}\label{subsec:uarchreview}
\begin{figure}[ht]
  \centering
  \includegraphics[width=\columnwidth]{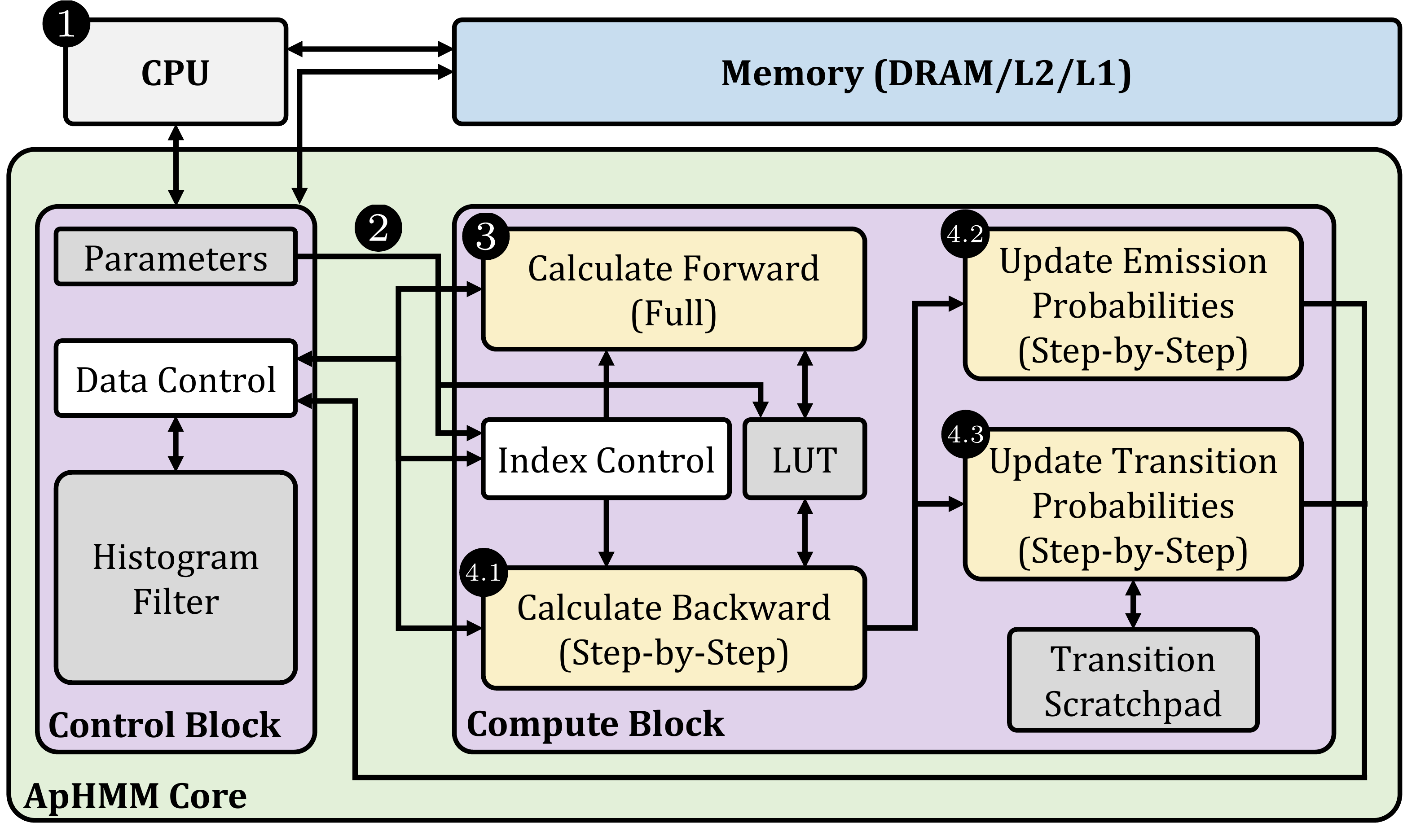}
  \caption{Overview of \proposal}
  \label{fig:aphmm-overview}
\end{figure}

\proposal provides a \textbf{flexible}, high-performant, and energy-efficient hardware-software co-designed acceleration framework for calculating each step in the Baum-Welch algorithm. Figure~\ref{fig:aphmm-overview} shows the main flow of \proposal when executing the Baum-Welch algorithm for pHMMs. To exploit the massive parallelism that DNA and protein sequences provide, \proposal processes many sequences in parallel using multiple copies of hardware units called \emph{\proposal Cores}. Each \proposal core aims to accelerate the Baum-Welch algorithm for pHMMs. An \proposal core contains two main blocks: 1)~\contb and 2)~\compb. \contb provides efficient on- and off-chip synchronization and communication with CPU, DRAM, and L2/L1 cache. \compb efficiently and effectively performs each step in the Baum-Welch algorithm: 1)~Forward calculation, 2)~Backward calculation, and 3)~Parameter Updates with respect to their corresponding equations in Section~\ref{subsec:BW_alg}.

\proposal starts when the CPU loads necessary data to memory and sends the parameters to \proposal \circlednumber{$1$}. \proposal uses the parameters to decide on the pHMM design (i.e., either traditional pHMM design or modified design for error correction) and steps to execute in the Baum-Welch algorithm. The parameters related to design are sent to \compb so that each \compb can efficiently make proper state connections \circlednumber{$2$}. For each character in the input sequence that we aim to calculate the similarity score, \compb performs 1)~Forward, 2)~Backward, 3)~and Parameter Updates steps. \proposal enables disabling the calculation of Backward and Parameter Updates steps if they are not needed for an application. \proposal iterates over the entire input sequence to fully perform the Forward calculation with respect to Equation~\ref{eq:forward}\circlednumber{$3$}. \proposal then re-iterates each character on the input sequence character-by-character to perform the Backward calculations for each timestamp $t$ with respect to Equation~\ref{eq:backward} (i.e., step-by-step) \circlednumberb{$4.1$}. \proposal updates emission \circlednumberb{$4.2$} and transition probabilities \circlednumberb{$4.3$} as the Backward values are calculated in each timestamp.

\subsection{Control Block}\label{subsec:control}
\contb is responsible for managing the input and output flow of the compute section efficiently and correctly by issuing both memory requests and proper commands to \compb to configure for the next set of operations (e.g., the forward calculation for the next character of the sequence $S$). Figure~\ref{fig:aphmm-overview} shows three main units in \contb: 1)~\pars, 2)~\datacont, and 3)~\hif.

\head{\pars} \contb contains the parameters of pHMM and the Baum-Welch algorithm. These parameters define 1)~pHMM design (i.e., either the traditional design or modified design for error correction) and 2)~steps to execute in the Baum-Welch algorithm as \proposal allows disabling the calculation of Backward or Parameter Updates steps.

\head{\datacont} To ensure the correct, efficient, and synchronized data flow, \proposal uses \datacont to 1) arbitrate among the read and write clients and 2) pipeline the read and write requests to the memory and other units in the accelerator (e.g., \hif). Data control is the main memory management unit for issuing a read request to L1 cache to obtain 1)~each input sequence $S$, 2)~corresponding pHMM graph (i.e., $G(V,A)$), 3) corresponding parameters and coefficients from the previous \emph{timestamp} (e.g., Forward coefficients from timestamp $t-1$ as shown in Equation~\ref{eq:forward}). \datacont collects and controls the write requests from various clients to ensure data is synchronized.

\head{\hif} The filtering approach is beneficial for eliminating negligible states from Forward and Backward calculations without significantly compromising accuracy (Section~\ref{sec:motivation}). The \textbf{challenge} in implementing a straightforward filtering mechanism lies in performing sorting in hardware, which is difficult to achieve efficiently. Our \textbf{key idea} is to replace the sorting mechanism with a histogram-based filter, allowing values to be placed into different bins based on their Forward or Backward values. This offers quick and approximate identification of non-negligible states (i.e., states with the best values until the filter is full) based on their bin locations. To enable such a binning mechanism, we employ a \textbf{flexible} \emph{histogram-based} filtering mechanism in the \proposal on-chip memory.

\begin{figure*}[ht]
  \centering
  \includegraphics[width=0.7\linewidth]{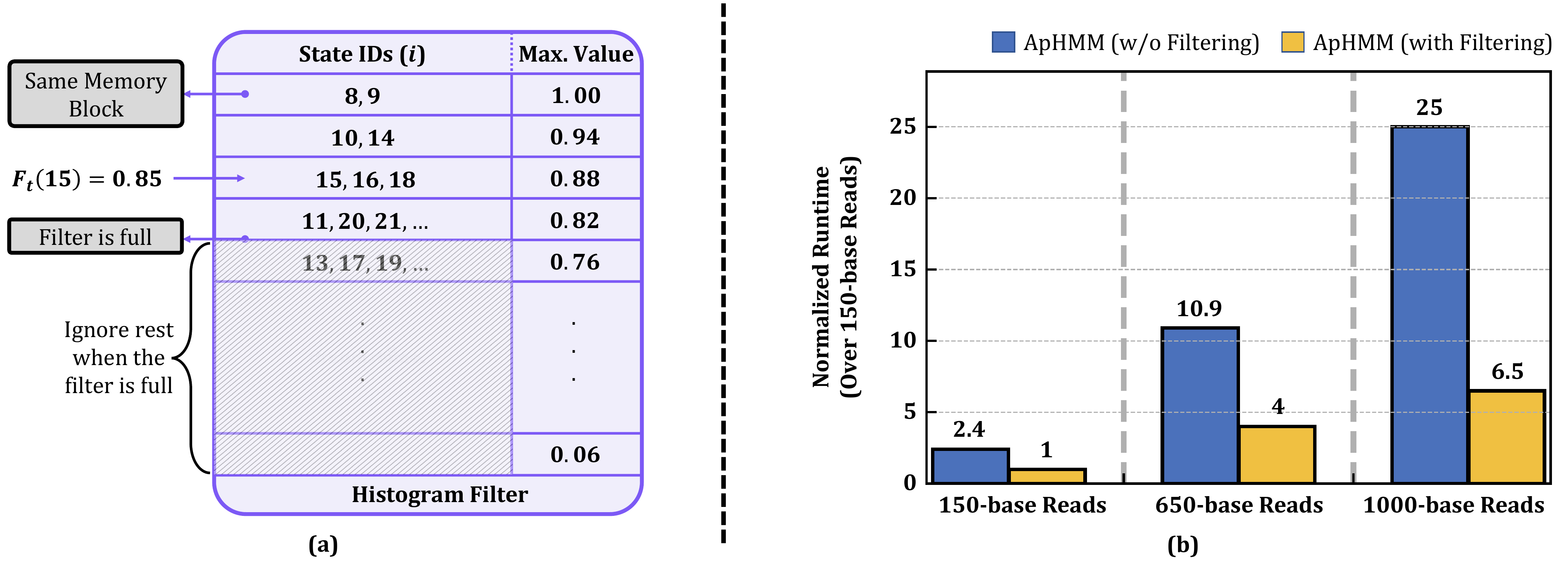}
    \caption{(a) Overall structure of a \hif. (b) Effect of the \hif approach in \proposal for different sequence lengths.}
    \label{fig:filtering-histogram}
\end{figure*}

Figure~\ref{fig:filtering-histogram}(a) shows the overall structure of our \hif. The filter categorizes states into bins based on their Forward or Backward values at the current execution timestamp in three steps. First, \hif divides the $[0, 1]$ range into $n$ bins, with each bin corresponding to a specific range of Forward or Backward values. The range for each bin is $1/n$. We empirically chose 16 bins, ensuring a range of $1/16 = 0.0625$, to maintain the same accuracy when the filter size is 500 (Figure~\ref{fig:motivation-filtersize}). For simplicity, we use $0.06$ as the range value in Figure~\ref{fig:filtering-histogram}(a), with the maximum value in each bin's range displayed under \emph{Max. Value}.

Second, the \hif assigns addresses to states such that all states with Forward or Backward values within the same range fall into the same memory block. This addressing mechanism employs a \emph{base and offset} strategy, where the base represents the start of the memory block for a specific range of values, and the offset is the pointer to the next free memory region within the memory block. This base and offset strategy allows \proposal to discard negligible states efficiently, as their addresses are known without sorting.

Third, to identify the addresses of negligible states, the \hif accumulates the count of states in each bin, starting with the bin with the largest \emph{Max. Value} (i.e., $1.00$). When the overall state count exceeds the filter size (e.g., 500), the remaining bins are assumed to contain only negligible states. The \hif can find all the non-negligible states that a filtering technique with a sorting mechanism finds, albeit with the cost of including states beyond the predetermined filter size, as the accumulated state count in the last bin can exceed the filter count. While it is possible to perform additional computations in the last bin to prevent exceeding the filter size, we leave such optimization for future work.

To build a \textbf{flexible} framework for various applications, The microarchitecture is configurable to vary the number of bins ($n$) based on the application and the average sequence length. We recommend conducting an empirical analysis before determining this range for a particular application, as it may vary significantly.

\proposal offers the option to disable the filtering mechanism if the application does not necessitate a filter operation for more optimal computations. As shown in Figure~\ref{fig:filtering-histogram}(b), there is a trade-off when using \proposal with and without a filter for sequences of varying lengths. We observe that the performance significantly improves when the filtering mechanism is enabled, especially for longer sequences. This can be primarily attributed to the fact that the number of states requiring processing at the subsequent timestamp can exponentially increase, as each state typically has more than one transition, potentially leading to an exponential increase in states at each subsequent timestamp. As the sequence length grows, such an exponential increase can adversely affect the application, which can be significantly mitigated without compromising accuracy through a filtering approach.

\subsection{Compute Block}\label{subsec:computeblock}
\begin{figure}[ht]
  \centering
  \includegraphics[width=\columnwidth]{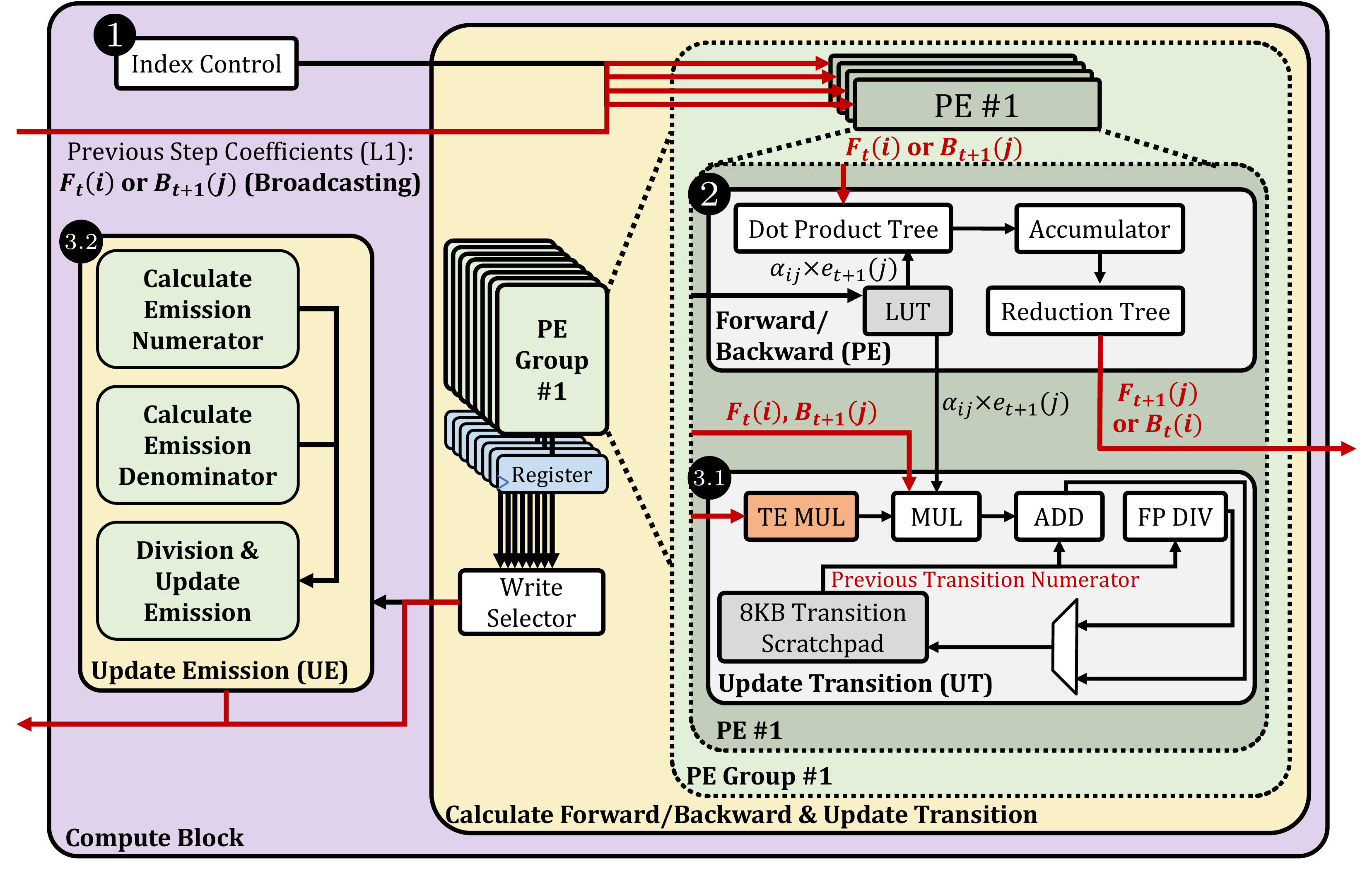}
  \caption{Overview of a \compb. Red arrows show on- and off-chip memory requests.}
  \label{fig:aphmm-compute}
\end{figure}

Figure~\ref{fig:aphmm-compute} shows the overall structure of a \compb, which is responsible for performing core compute operations of each step in the Baum-Welch algorithm (Figure~\ref{fig:aphmm-overview}) based on the configuration set by the \contb via Index Control\circlednumber{$1$}. A \compb contains two major units: 1)~a unit for calculating Forward (Equation~\ref{eq:forward}) and Backward (Equation~\ref{eq:backward}) values\circlednumber{$2$} and updating transition probabilities (Equation~\ref{eq:transition}) \circlednumberb{$3.1$}, and 2)~a unit for updating the emission probabilities (Equation~\ref{eq:emission}) \circlednumberb{$3.2$}. Each unit performs the corresponding calculations in the Baum-Welch algorithm.

\head{Forward and Backward Calculations}
Our goal is to calculate the forward and backward values for all states in a pHMM graph $G(V,A)$, as shown in Equations~\ref{eq:forward} and ~\ref{eq:backward}, respectively. To calculate the Forward or Backward value of a state $i$ at a timestamp $t$, \proposal uses Processing Engines (\pes). Since pHMMs may require processing hundreds to thousands of states to process at a time, \proposal includes many \pes and groups them \pegs. Each \pe is responsible for calculating the Forward and Backward values of a state $v_i$ per timestamp $t$. Our \textbf{key challenge} is to balance the utilization of the compute units with available memory bandwidth. We discuss this trade-off between the number of \pes and memory bandwidth in Section~\ref{sec:hw-config}. To efficiently calculate the Forward and Backward values, \pe performs two main operations.

First, \pe uses the parallel four lanes in \dotpt and \accum to perform multiple multiply and accumulation operations in parallel, where the final summation is calculated in the \redt. This design enables efficient multiplication and summation of values from previous timestamps (i.e., $F_{t-1}(j)$ or $B_{t+1}(j)$). Second, to avoid redundant multiplications of transition and emission probabilities, \textbf{the key idea} in \pes is to efficiently enable the reuse of the products of these common multiplications. To achieve this, our key mechanism stores these common products in lookup tables (\luts) in each \pe while enabling efficient retrievals of the common products. We store these products as these values can be preset (i.e., fixed) before the training step starts and frequently used during training while causing high computational overheads.

Our \textbf{key challenge} is to design minimal but effective \luts to avoid area and energy overheads associated with \luts without compromising the computational efficiency \luts provide. To this end, we analyze error correction, protein family search, and multiple sequence alignment implementations. We observe that 1) redundant multiplications are frequent only during training and 2) the alphabet size of the biological sequence significantly determines the number of common products (i.e., 4 in DNA and 20 in proteins). Since error correction is mainly bottlenecked during the training step, we focus on the DNA alphabet and the pHMM design that error correction uses. We identify that each state uses 1) at most $4$ different emission probabilities (i.e., DNA letters) and 2) on average $7$ different transitions. This results in $28$ different combinations of emission and transition probabilities. To enable slightly better flexibility, we assume $9$ different transitions and include 36 entries in \luts.

The \textbf{key benefit} is \luts provide \proposal with a bandwidth reduction of up to $66\%$ per \pe while avoiding redundant computations. \proposal is \textbf{flexible} such that it enables disabling the use of \luts and instead performing the actual multiplication of transition and emission probabilities (\temul unit in Figure~\ref{fig:aphmm-compute}).

\head{Updating the Transition Probabilities}
Our goal is to update the transition probabilities of all the states, as shown in Equation~\ref{eq:transition}. To achieve this, we design the \emph{Update Transition} (\ut) compute unit and tightly couple it with \pes, as shown in Figure~\ref{fig:aphmm-compute}. Each \ut efficiently calculates the denominator and numerator in Equation~\ref{eq:transition} for a state $v_i$. \uts include three key mechanisms.

First, to enable efficient broadcasting of common values between the Backward calculation and Parameter Updates steps, \proposal connects \pes with \uts for updating transitions. Each \pe in a \peg is broadcasted with the \emph{same} previously calculated $F_{t}(i)$ or $B_{t+1}(j)$ values from the previous timestamp for calculating the $F_{t+1}(j)$ or $B_{t}(i)$ values, respectively. The incoming red arrows in Figure~\ref{fig:aphmm-compute} show the flow of these Forward and Backward values in \pes and \uts. The calculation of $F_{t+1}(j)$ involves a summation over all states $i$ as shown in Equation~\ref{eq:forward}. The $F_{t}(i)$ term is common to the calculation of $F_{t+1}(j)$ for all states $j$ and hence can be broadcasted. Similarly, the calculation of $B_{t}(i)$ involves a summation over all states $j$ (Equation~\ref{eq:backward}). The $B_{t+1}(j)$ term is common to the calculation of $B_{t}(i)$ for all states $i$ and hence can be broadcasted. This \textbf{key design choice} exploits the broadcast opportunities available within the common multiplications in the Baum-Welch equations.

Second, \proposal cores are designed to directly consume the broadcasted Backward values in when updating the emission and transition probabilities to reduce the bandwidth and storage requirements. We exploit the broadcasting opportunities because we observe that Backward values do \emph{not} need to be fully computed, and they can be directly consumed when updating the transitions and emission probabilities while the Backward values are broadcasted in the current timestamp. \proposal updates emission and transition probabilities step-by-step as Backward values are calculated, a hardware-software optimization we call the \emph{partial compute approach}. It is worth noting that \proposal fully computes and stores the Forward values before updating the emission and transition probabilities. \textbf{The key benefits} of combining broadcasting with the partial compute approach are 1)~decoupling hardware scaling from bandwidth requirements and 2)~reducing the bandwidth requirement by $4\times$ (i.e., 32 bits/cycle instead of 128 bits/cycle).

Third, to exploit the spatiotemporal locality in pHMMs, we utilize on-chip memory in \uts with memoization techniques that allow us to store the recent transition calculations. We observe from Equation~\ref{eq:transition} that transition update is calculated using the values of states connected to each other. Since the connections are predefined and provide spatial locality (Figure~\ref{fig:motivation-datadependency}), our \textbf{key idea} is to memoize the calculation of all the numerators from the same $i$ to different states by storing these numerators in the same memory space. This enables us to process the same state $i$ in different timestamps within the same \pe to reduce the data movement overhead within \proposal. To this end, we use an 8KB on-chip memory (Transition Scratchpad) to store and reuse the result of the numerator of Equation~\ref{eq:transition}. Since we store the numerators that contribute to all the transitions of a state $i$ within the same memory space, we perform the final division in Equation~\ref{eq:transition} by using the values in the Transition Scratchpad. We use an 8KB memory as this enables us to store 256 different numerators from any state $i$ to any other state $j$. We observe that pHMMs have 3-12 distinct transitions per state. Thus, 8KB storage enables us to operate on at least 20 different states within the same \pe. \textbf{The memoization technique allows} 1) skipping redundant data movement and 2) reducing the bandwidth requirement by $2 \times$ per \ut.

\head{Updating the Emission Probabilities}
Our goal is to update the emission probabilities of all the states, as shown in Equation~\ref{eq:emission}. To achieve this, we use the \emph{Update Emission (\ue)} unit, as shown in Figure~\ref{fig:aphmm-compute}, which includes three smaller units: 1)~Calculate Emission Numerator, 2)~Calculate Emission Denominator, and 3)~Division \& Update Emission. \ue performs the numerator and denominator computations in parallel as they are independent of each other, which includes a summation of the products $F_{t}(i)B_t(i)$.
These $F_{t}(i)$ and $B_t(i)$ values are used to update \emph{both} the transition and emission probabilities, as shown in Equation~\ref{eq:transition}.
To reduce redundant computations, our \textbf{key design} choice is to use the $F_{t}(i)$ and $B_t(i)$ values as broadcasted in the transition update step since these values are also used for updating the emission probabilities. Thus, we broadcast these values to \ues through \emph{Write Selectors}, as shown in Figure~\ref{fig:aphmm-compute}.

An \proposal core writes and reads both the numerator and denominator values to L1 cache to update the emission probabilities. The results of the division operations and the posterior emission probabilities (i.e., $e^{*}_{X}(v_{i})$ in Equation~\ref{eq:emission}) are written back to L1 cache after processing each read sequence $S$. If we assume that the number of characters in an alphabet $\Sigma$ is $n_{\Sigma}$ (e.g., $n_{\Sigma} = 4$ for DNA letters), \proposal stores $n_{\Sigma}$ many different numerators for each state of the graph as emission probability may differ per character for each state. Our microarchitecture design is \textbf{flexible} such that it allows defining $n_{\Sigma}$ as a parameter.

\subsection{Hardware Configuration Choice}\label{sec:hw-config}
Our goal is to identify the ideal number of memory ports and processing elements (\pe) for better scaling \proposal with many cores. We identify the number of memory ports and their dependency on the hardware scaling in four steps.
First, \proposal requires one input memory port for reading the input sequence to update the probabilities in a pHMM graph.
Second, updating the transition probabilities requires $3$ memory ports: 1)~reading the forward value from L1, 2)~reading the transition, and 3)~emission probabilities if using the \luts is disabled (Section~\ref{subsec:computeblock}). Since these ports are shared across each \pe, the number of \pes and memory bandwidth per port determines the utilization of these memory ports.
Third, \proposal requires $4$ memory ports to update the emission probabilities for 1)~calculating the numerator and 2)~denominator in Equation~\ref{eq:emission}, 3)~reading the forward from Write Selectors, and 4)~writing the output. These memory ports are \emph{independent} of the impact of the number of \pes in a single \proposal core.
Fourth, \proposal does not require additional memory ports for each step in the Baum-Welch algorithm due to the broadcasting feature of \proposal (Section~\ref{subsec:computeblock}). Instead, computing these steps depends on 1)~memory bandwidth per port, which determines the number of multiplications and accumulations in parallel in a \pe, and 2)~ the number of processing engines (\pes). 
We conclude that the overall requirement for a \proposal core is $8$ memory ports with the same bandwidth per port.

\begin{figure}[ht]
  \centering
  \includegraphics[width=\columnwidth]{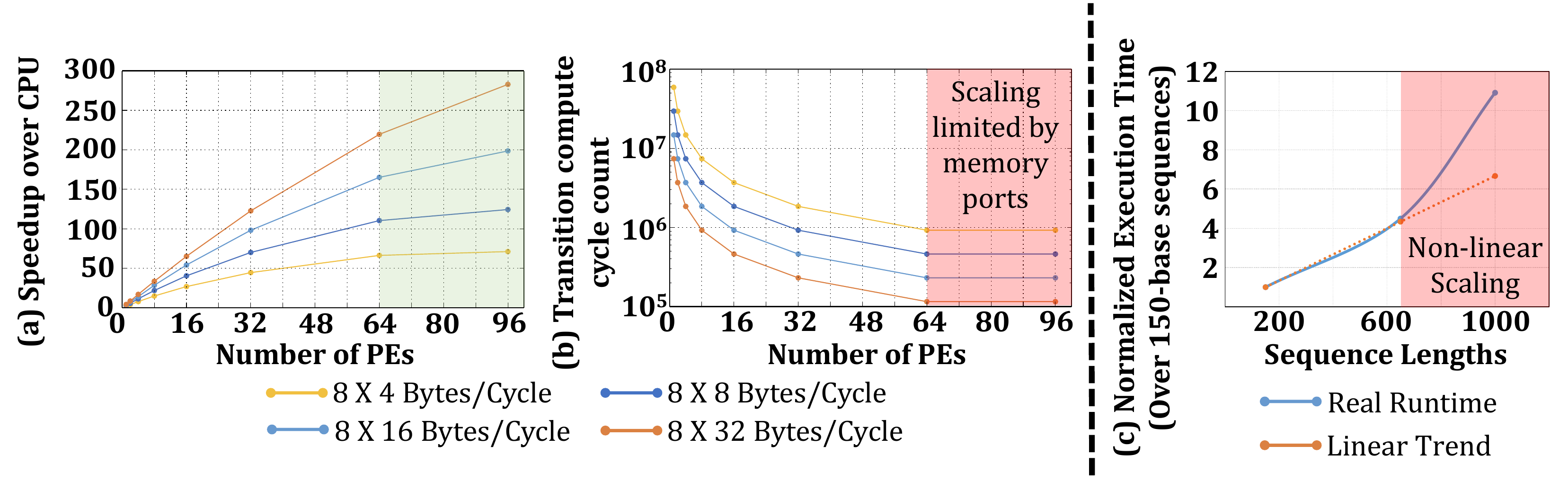}
  \caption{(a) Acceleration scaling with the number of \pes. (b) Compute cycle acceleration when calculating the transition probabilities with the increased number of \pes. (c) Increase in the execution time with respect to the sequence (chunk) lengths. The data points are for sequence lengths 150, 650, and 1000, respectively. The linear trend shows the expected linear increase in execution time, and the real runtime shows the actual runtimes.}
  \label{fig:hw_scaling}
\end{figure}

Figure~\ref{fig:hw_scaling} shows the scaling capabilities of \proposal with the number of \pes and sequence length to decide 1)~the overall number of \pes and 2)~the longest chunk size for the best acceleration. First, to decide the overall number of \pes to use in \proposal, in Figure~\ref{fig:hw_scaling}(a), we show the acceleration speedup while scaling \proposal with the number of \pes and bandwidth per memory port, where we keep the number of memory ports fixed to $8$. We observe that a linear trend of increase in acceleration is possible until the number of \pes reaches $64$, where the rate of acceleration starts reducing. We explore the reason for such a trend in Figure~\ref{fig:hw_scaling}(b). We find that the acceleration on the transition step starts settling down as the number of \pes grows due to memory port limitation that reduces parallel data read from memory per \pe, eventually resulting in the underutilization of resources. Second,  We conclude that the acceleration trend we observe in Figure~\ref{fig:hw_scaling}(a) is mainly due to the scaling impact on the forward and backward calculation when the number of \pes is greater than $64$ where $8$ memory ports start becoming the bottleneck.

In our design, the choice of memory bandwidth influences the number of \pegs and \pes, given a constant number of memory ports. While our hardware can scale to accommodate higher bandwidths, we opt for a 16-bytes/cycle bandwidth. This design choice aligns with the 128-bit line size of our L1 cache, allowing us to operate on four single-precision floating-point values (32-bit) across 4 \pes simultaneously. To fully utilize all 64 \pes, as discussed earlier, we employ 16 \pegs (64 \pes = 4 \pes $\times$ 16 \pegs).

Second, to identify the optimal chunk size (i.e., sequence length) that ensures a near-linear increase in execution time with increasing sequence length, we examine the execution time of the Baum-Welch algorithm for chunk sizes of 150, 650, and 1000 bases, as shown in Figure~\ref{fig:hw_scaling}(c). We observe a linear increase in execution time with chunk sizes up to approximately 650 bases. Beyond this point, the execution time begins to increase non-linearly. This non-linear scaling is primarily due to the increased cache space requirements for storing certain parameters (e.g., emission values), as shown in Supplemental Figure S1. This increased cache pressure leads to more data accesses from the upper levels of the memory hierarchy. \proposal can maintain a linear trend in execution time for longer sequences by either increasing the L1 and L2 cache capacities or utilizing higher-bandwidth memories to mitigate the data movement overheads. We provide further details regarding the data distribution and memory layout in Supplemental Section S2.

We conclude that the memory ports and chunk size primarily constrain the acceleration speedup of \proposal, as the \pes start to be underutilized due to increased data movement overheads. To further enhance the acceleration with \proposal, optimizing the utilization of \pes by minimizing these overheads is crucial.

\begin{figure*}[hbt]
\centering
\includegraphics[width=0.8\linewidth]{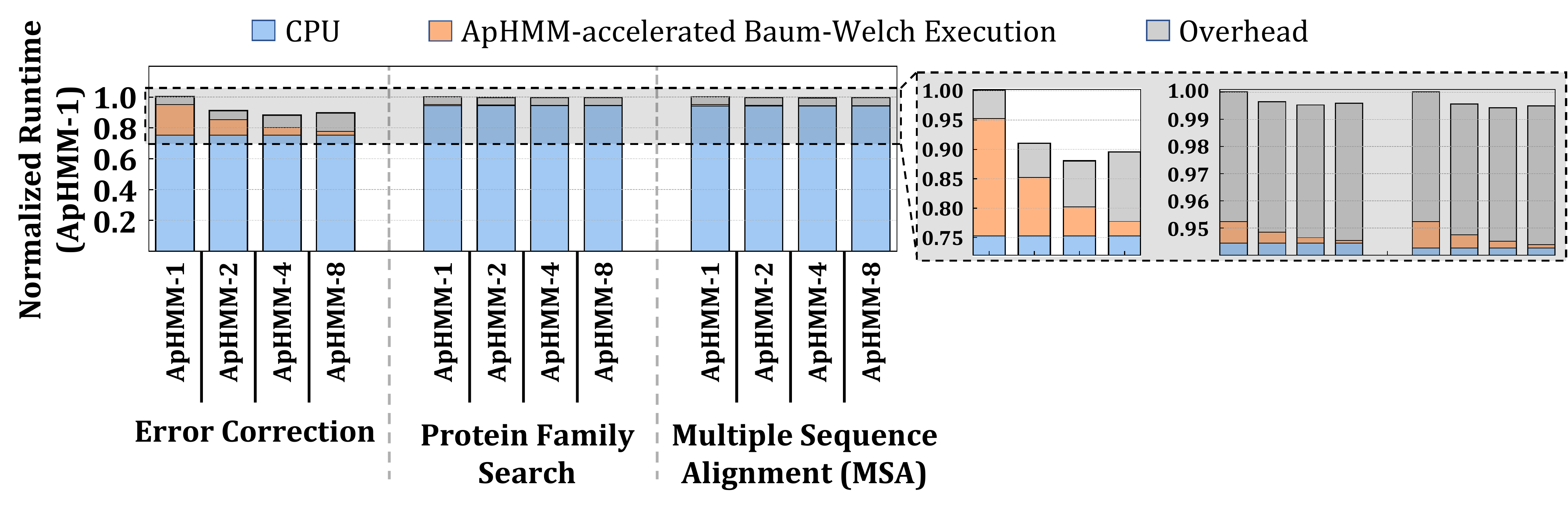}
\caption{Normalized runtimes of multi-core \proposal compared to the single-core \proposal (\proposal-1).}
\label{fig:aphmm_cores}
\end{figure*}

\head{Number of \proposal Cores}
We show our methodology for choosing the ideal number of cores in \proposal for accelerating the applications. Figure~\ref{fig:aphmm_cores} shows the speedup of three bioinformatics applications when using single, 2, 4, and 8 cores in \proposal. We divide the entire execution time of the applications into three stages: 1)~the CPU execution of the application that does not use the Baum-Welch execution, 2)~the Baum-Welch execution accelerated using \proposal, and 3)~and the overhead mainly caused due to data movements. Our analysis incorporates the estimated off- and on-chip data movement overhead. We observe that using 4 cores in \proposal provides the best speedup overall. This is because the applications provide smaller rooms for acceleration for two reasons. First, the remaining CPU part of the application becomes the bottleneck in the entire execution of the application due to the significant acceleration of the Baum-Welch execution using \proposal. Second, the data movement overhead starts causing more latency than the benefits of further accelerating the Baum-Welch algorithm by increasing the number of cores. This suggests \proposal is bounded by the data movement overhead when scaling it to a larger number of cores, and there is still room for improving the performance of \proposal by placing \proposal inside or near the memory (e.g., high-bandwidth memories) to reduce the data movement overheads that limit scaling \proposal to many cores. Based on our observations, we use a 4-core \proposal to achieve the best overall performance (see Supplemental Section S3 for the execution flow of the system with multiple cores in \proposal).

%% file: sections/5_evaluation.tex
\section{Evaluation} \label{sec:results}
We evaluate our acceleration framework, \proposal, for three use cases: 1)~error correction, 2)~protein family search, and 3)~multiple sequence alignment (MSA). We compare our results with the CPU, GPU, and FPGA implementations of the use cases.

\subsection{Evaluation Methodology}\label{subsec:evalmethod}
We use the configurations shown in Table~\ref{tab:ApHMM_parameters} to implement the \proposal design described in Section~\ref{sec:aphmm} in SystemVerilog. We carry out synthesis using Synopsys Design Compiler~\cite{noauthor_tool_nodate} in a typical 28nm process technology node at 1GHz clock frequency with tightly integrated on-chip memory (1GHz) to extract the logic area and power numbers. We develop an analytical model to extract performance and area numbers for a scale-up configuration of \proposal. We use 4 \proposal cores in our evaluation (Section~\ref{sec:hw-config}). We account for an additional 5\% of cycles to compensate for the arbitration across memory ports. These extra cycles estimate the cycles for synchronously loading data from DRAM to L2 memory of a single \proposal core and asynchronous accesses to DRAM when more data needs to be from DRAM for a core (e.g., Forward calculation may not fit the L2 memory).

\begin{table}[ht]
\centering
\caption{Microarchitecture Configuration}\label{tab:ApHMM_parameters}
\resizebox{\columnwidth}{!}{
\begin{tabular}{@{}ll@{}}\toprule
Memory & Memory BW (Bytes/cycle): 16, Memory Ports (\#): 8\\
& L1 Cache Size: 128KB \\\cmidrule{1-2}
Processing & \pes (\#): 64,  Multipliers per \pe (\#): 4, Adders per \pe (\#): 4\\
Engine & Memory per PE: 8, Update Transitions (\#): 64, Update Emissions (\#): 4 \\ \bottomrule
\end{tabular}
}
\end{table}

We use the CUDA library~\cite{nickolls_scalable_2008} (version 11.6) to provide a GPU implementation of the software optimizations described in Section~\ref{sec:aphmm} for executing the Baum-Welch algorithm. Our GPU implementation, \textbf{\proposal-GPU}, uses the pHMM design designed for error correction, implements \luts (Section~\ref{subsec:computeblock}) as shared memory, and uses buffers to arbitrate between current and previous Forward/Backward calculations to reflect the software optimizations of \proposal in GPUs. We integrate our GPU implementation with a pHMM-based error correction tool, Apollo~\cite{firtina_apollo_2020}, to evaluate the GPU implementation. Our GPU implementation is the first GPU implementation of the Baum-Welch algorithm for profile Hidden Markov models.

We use gprof~\cite{graham_gprof_2004} to profile the baseline CPU implementations of the use cases on the AMD EPYC 7742 processor (2.26GHz, 7nm process) with single- and multi-threaded settings. We use the CUDA library and \emph{nvidia-smi} to capture the runtime and power usage of \proposal-GPU on NVIDIA A100 and NVIDIA Titan V GPUs, respectively.

We compare \proposal with the CPU, GPU, and FPGA implementations of the Baum-Welch algorithm and use cases in terms of execution time and energy consumption. To evaluate the Baum-Welch algorithm, we execute the algorithm in Apollo~\cite{firtina_apollo_2020} and calculate the average execution time and energy consumption of a single execution of the Baum-Welch algorithm. To evaluate the end-to-end execution time and energy consumption of error correction, protein family search, and multiple sequence alignment, we use Apollo~\cite{firtina_apollo_2020}, hmmsearch~\cite{eddy_accelerated_2011}, and hmmalign~\cite{eddy_accelerated_2011}. We replace their implementation of the Baum-Welch algorithm with \proposal when collecting the results of the end-to-end executions of the use cases accelerated using \proposal. When available, we compare the use cases that we accelerate using \proposal to their corresponding CPU, GPU, and FPGA implementations. For the GPU implementations, we use both \proposal-GPU and HMM\_cuda~\cite{yu_gpu-accelerated_2014}. For the FPGA implementation, we use the FPGA Divide and Conquer (D\&C) accelerator proposed for the Baum-Welch algorithm~\cite{pietras_fpga_2017}. When evaluating the FPGA accelerator, we ignore the data movement overhead and estimate the acceleration based on the speedup provided by the earlier work. We acknowledge that the performance and energy comparisons can be attributed to both platform differences and architectural optimizations, especially when comparing \proposal with the FPGA accelerator. Although our evaluations lack comparisons in the equivalent platforms for FPGAs, we still believe that our evaluations provide valuable insights regarding the benefits of our ASIC implementation compared to the FPGA work.

In terms of accuracy, we ensure the accuracy of our results by faithfully implementing all the equations of the Baum-Welch algorithm and rigorously testing their output during our ASIC design. The only exception is the \hif, where we introduce a binning approach to include all the states a sorting-based software implementation would include, ensuring at least the same minimum accuracy as the original software implementation. Our accuracy evaluation shows that the histogram filter approach usually leads to better accuracy than the sorting approach, with a negligible accuracy difference within a +-0.2\% range. To reproduce the output for comparison purposes, we provide the source code of our software optimizations in the GPU implementation of \proposal (\proposal-GPU).

\head{Data Set}
To evaluate the error correction use case, we prepare the input data that Apollo requires: 1) assembly and 2) read mapping to the assembly. To construct the assembly and map reads to the assembly, we use reads from a real sample that includes overall 163,482 reads of Escherichia coli (\emph{E. coli}) genome sequenced using PacBio sequencing technology with the average read length of 5,128 bases. The accession code of this sample is SAMN06173305. Out of 163,482 reads, we randomly select 10,000 sequencing reads for our evaluation. We use minimap2~\cite{li_minimap2_2018} and miniasm~\cite{li_minimap_2016} to 1) find overlapping reads and 2) construct the assembly from these overlapping reads, respectively. To find the read mappings to the assembly, we use minimap2 to map the same reads to the assembly that we generate using these reads. We provide these inputs to Apollo for correcting errors in the assembly we construct.

To evaluate the protein family search, we use the protein sequences from a commonly studied protein family, Mitochondrial carrier (PF00153), which includes 214,393 sequences with an average length of 94.2. We use these sequences to search for similar protein families from the entire Pfam database~\cite{mistry_pfam_2021} that includes 19,632 pHMMs. To achieve this, the hmmsearch~\cite{eddy_accelerated_2011} tool performs the Forward and Backward calculations to find similarities between pHMMs and sequences.

To evaluate the multiple sequence alignment, we use 1,140,478 protein sequences from protein families Mitochondrial carrier (PF00153), Zinc finger (PF00096), bacterial binding protein-dependent transport systems (PF00528), and ATP-binding cassette transporter (PF00005). We align these sequences to the pHMM graph of the Mitochondrial carrier protein family. To achieve this, the hmmalign~\cite{eddy_accelerated_2011} tool performs the Forward and Backward calculations to find similarities between a single pHMM graph and sequences.

\textbf{Data Set Justification}
In our study, we carefully chose our datasets for overhead analysis and evaluation. We believe these datasets are comprehensive and relevant to guide our ASIC design and to evaluate \proposal with other systems for several reasons. First, our datasets cover various use cases with various sequence lengths (i.e., an average read length of 5,168 and an average protein sequence length of 94.2) and alphabet sizes (4 in DNA and 20 in proteins). This diversity ensures that our results are not skewed toward a specific use case or dataset.

Second, for error correction, we use a real-world sample of the \emph{E. coli} genome, a commonly studied bacterial genome. The overall length of randomly selected 10,000 \emph{E. coli} reads is around 50,000,000 bases (the average length of a single read is 5,168). This ensures that these reads cover the entire \emph{E. coli} genome around 10 times (i.e., $10\times$ depth of coverage), which ensures that the Baum-Welch algorithm is executed by performing error correction on the entire genome multiple times without focusing on the specific regions of the genome to avoid potential bias that can be caused on particular regions. For the multiple sequence alignment and the protein family search, we use the most commonly studied protein families as these protein families are among the top 20 families with the largest number of protein sequence alignments\footnote{Top 20 protein families can be found at \url{http://pfam-legacy.xfam.org/family/browse?browse=top\%20twenty}}, ensuring the relevance and applicability of our work. The bottleneck analysis was conducted on a subset of these datasets, demonstrating that our design is not overfitting to a specific dataset.

Third, the Baum-Welch algorithm operates on a sub-region of the pHMM graph, the size of which is determined by the sequence length or chunk size, whichever is shorter. Thus, the complexity of a single Baum-Welch execution on this sub-region is determined mainly based on the sequence length and the alphabet size, regardless of the overall genome size or the sequence lengths larger than the chunk size. In our case, we cover all these cases: 1) the pHMM subgraph is determined based on the sequence length (around 90 bases) as it is shorter than the chunk size (up to 1000 bases) in the protein family search and the multiple sequence alignment 2) the length of the pHMM subgraph is determined by the chunk size in error correction as the sequence length is usually larger (around 10,000 bases) than the chunk size, and 3) different alphabet sizes in DNA and protein.

Fourth, for overhead analysis, we discuss in Section~\ref{subsec:inefficiencies}, we ensure our design is not overfitting to a specific dataset by using a subset of these datasets for each use case. The overhead was measured by taking the geometric mean across different runs to further ensure the robustness of our design. Since our ASIC design is mainly influenced based on the observations we make in our overhead analysis and to maximize the performance improvement for the applications mainly bottlenecked by the Baum-Welch algorithm (i.e., error correction), we believe the comprehensiveness of our data set choice and the overhead analysis enable us improving the robustness of our accelerator across a wide range of potentially many other use cases other than the use cases we evaluate in this work.

\subsection{Area and Power}\label{subsec:overhead}
Table~\ref{tab:ApHMM_area} shows the area breakup of the major modules in \proposal. For the area overhead, we find that the Update Transition (\ut) units take up most of the total area ($77.98\%$). This is mainly because \uts consist of several complex units, such as a multiplexer, division pipeline, and local memory. For the power consumption, \contb and PEs contribute to almost the entire power consumption ($86\%$) due to the frequent memory accesses these blocks make. Overall, a\proposal core incurs an area overhead of 6.5mm\textsuperscript{2} in 28nm with a power cost of 0.509W.

\begin{table}[ht]
\centering
\caption{Area and Power breakdown of \proposal}\label{tab:ApHMM_area}
\resizebox{0.8\columnwidth}{!}{
\begin{tabular}{@{}lrr@{}}\toprule
\textbf{Module Name} & \textbf{Area (mm$^2$)} & \textbf{Power (mW)} \\ \midrule
\contb & 0.011 & 134.4 \\
64 Processing Engines (\pes) & 1.333 & 304.2  \\
64 Update Transitions (\uts) & 5.097 & 0.8 \\
4 Update Emissions (\ues) & 0.094 & 70.4 \\
\textbf{Overall}  & 6.536 & 509.8 \\\cmidrule{1-3}
128KB L1-Memory & 0.632 & 100 \\ \bottomrule
\end{tabular}
}
\end{table}

\subsection{Accelerating the Baum-Welch Algorithm}\label{subsec:baumwelch}
\begin{figure}[htb]
  \centering
  \includegraphics[width=\columnwidth]{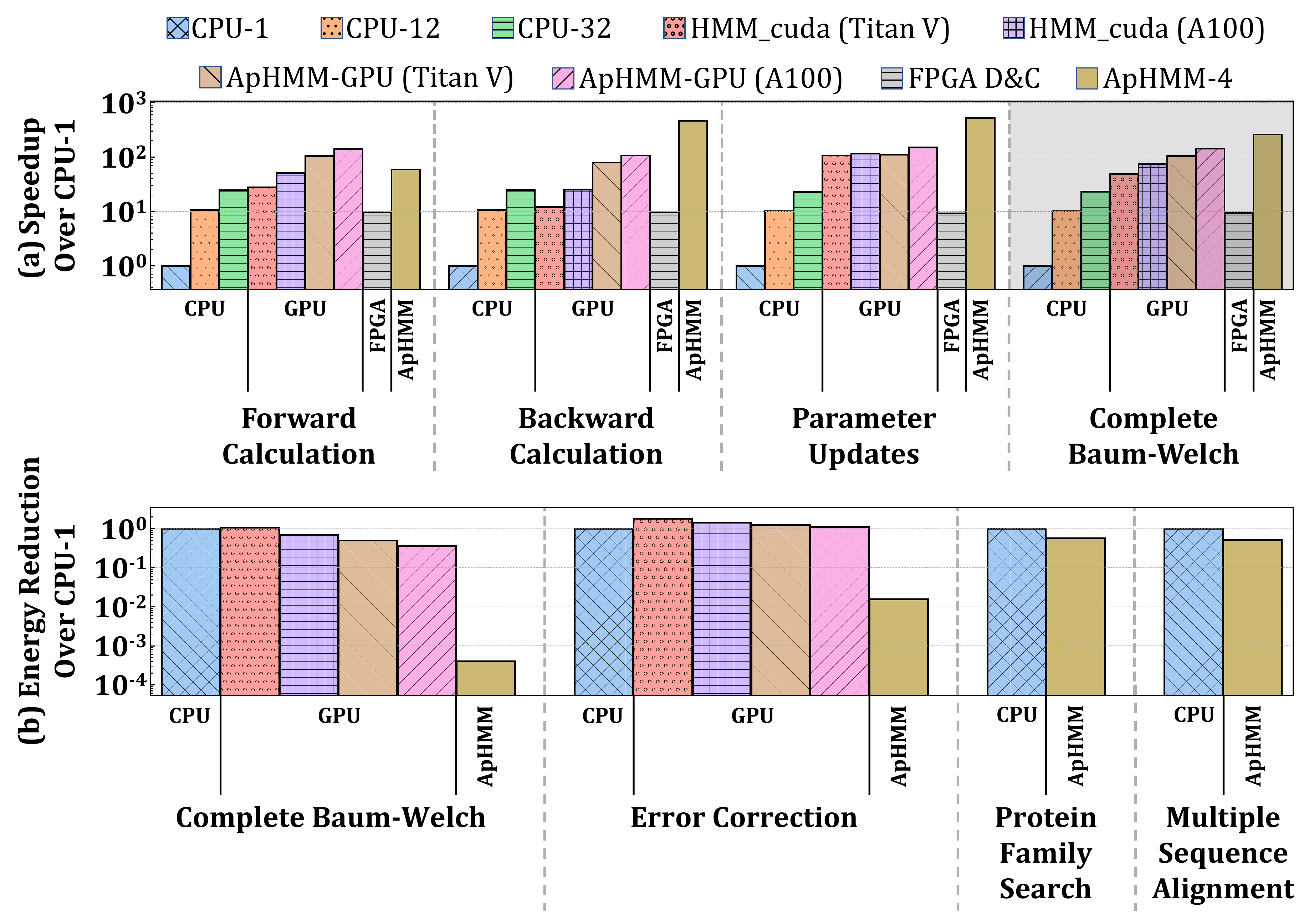}
    \caption{(a) Normalized speedups of each step in the Baum-Welch algorithm over single-threaded CPU (CPU-1). (b) Energy reductions compared to the CPU-1 implementation of the Baum-Welch algorithm and three pHMM-based applications.}
    \label{fig:baum_energy}
\end{figure}

Figure~\ref{fig:baum_energy} shows the performance and energy improvements of \proposal for executing the Baum-Welch algorithm. Based on these results, we make six key observations. 
First, we observe that \proposal is \pbaumcpuc - \pbaumcpua, \pbaumgpud-\pbaumgpua, and \pbaumfpga faster than the CPU, GPU, and FPGA implementations of the Baum-Welch algorithm, respectively. Although our evaluations do not directly compare the state-of-the-art FPGA work with the potential FPGA implementation of \proposal, we believe the performance benefits that \proposal provides arise not only from the differences in the platform and architecture but also from the optimizations we provide, which are absent in the existing FPGA work. We believe the benefits of these optimizations on the same platform can partly be observed when comparing \proposal-GPU with the state-of-the-art GPU accelerator.
Second, \proposal reduces the energy consumption for calculating the Baum-Welch algorithm by \ebaumcpua and \ebaumgpud-\ebaumgpua compared to the single-threaded CPU and GPU implementations, respectively. These speedups and reduction in energy consumption show the combined benefits of our software-hardware optimizations.
Third, the parameter update step is the most time-consuming step for the CPU and the GPU implementations, while \proposal takes the most time in the forward calculation step. The reason for such a trend shift is that \proposal reads and writes to L2 Cache and DRAM more frequently during the forward calculation than the other steps, as \proposal requires the forward calculation step to be fully completed and stored in the memory before moving to the next steps as we explain in Section~\ref{subsec:computeblock}. 
Fourth, we observe that \proposal-GPU performs better than HMM\_cuda by \pbaumgpucomp on average. HMM\_cuda executes the Baum-Welch algorithm on any type of hidden Markov model without a special focus on pHMMs. As we develop our optimizations based on pHMMs, \proposal-GPU can take advantage of these optimizations for more efficient execution.
Fifth, both \proposal-GPU and HMM\_cuda provide better performance for the Forward calculation than \proposal. We believe the GPU implementations are a better candidate for applications that execute only the Forward calculations as \proposal targets, providing the best performance for the complete Baum-Welch algorithm.
Sixth, the GPU implementations provide a limited speedup over the multi-threaded CPU implementations mainly because of frequent access to the host for synchronization and sorting (e.g., the filtering mechanism). These required accesses from GPU to host can be minimized with a specialized hardware design, as we propose in \proposal for performing the filtering mechanism.
We conclude that \proposal provides substantial improvements, especially when we combine speedups and energy reductions for executing the complete Baum-Welch algorithm compared to the CPU and GPU implementations, which makes it a better candidate to accelerate the applications that use the Baum-Welch algorithm than the CPU, GPU, and FPGA implementations.

\head{Breakdown of the optimizations benefits}
Table~\ref{tab:breakdown} shows the performance improvements that each \proposal optimization contributes for executing the Baum-Welch algorithm given the single-core hardware configuration we discuss in Section~\ref{sec:hw-config} compared to the CPU baseline of the Baum-Welch algorithm. We estimate the speedup of \hif by eliminating the sorting mechanism from filtering while considering the overhead of redundant states included in \hif. For other optimizations, we conservatively estimate the performance speedups by considering the memory bandwidth reductions that each optimization provides, as discussed in Section~\ref{sec:aphmm}, and the relation between acceleration speedup and the memory bandwidth requirements (Figure~\ref{fig:hw_scaling}).
We make five key observations. 
First, we find almost all optimizations contribute significantly to reducing the overall execution time of the Baum-Welch algorithm. Although \hif provides a limited speedup, this is because it constitutes around $8.5\%$ of the overall execution time (Observation 4 in Section~\ref{subsec:inefficiencies}). 
Second, the tight coupling of the broadcasting and the partial compute approach provides the most significant speedups by avoiding a large number of useless data movements with significant memory bandwidth reductions. 
Third, the speedup from \luts is mainly achieved by eliminating many single-precision floating-point operations, causing around $22.7\%$ of the total execution time (Observation 3 in Section~\ref{subsec:inefficiencies}). 
Fourth, the speedups with the memoization technique are purely achieved by significantly reducing the data movement latency when frequently calculating the transition probabilities.
Fifth, we find that the memoization and the partial compute optimizations are utilized only in the training step, and the \luts can be useful when the alphabet size is small (e.g., 4 in DNAs) due to storage limitations, which is usually the case when the Baum-Welch algorithm is used mainly for inference with the protein sequencing data. Although these benefits cannot be fully utilized in such cases, the remaining optimizations still provide a significant speedup up to $3.63\times$. We conclude that our optimizations achieve significant speedups for various use cases, from training with DNA sequencing data to inferring with protein sequencing data, allowing the acceleration of many applications that use the Baum-Welch algorithm with pHMMs.

\begin{table}[tbh]
\centering
\caption{Speedup of each Optimization over CPU.}
\resizebox{0.8\columnwidth}{!}{
\input{tables/breakdown}
}
\label{tab:breakdown}
\end{table}

\subsection{Use Case 1: Error Correction} \label{subsec:errorcorrection}
\begin{figure}[htb]
\centering
\includegraphics[width=\columnwidth]{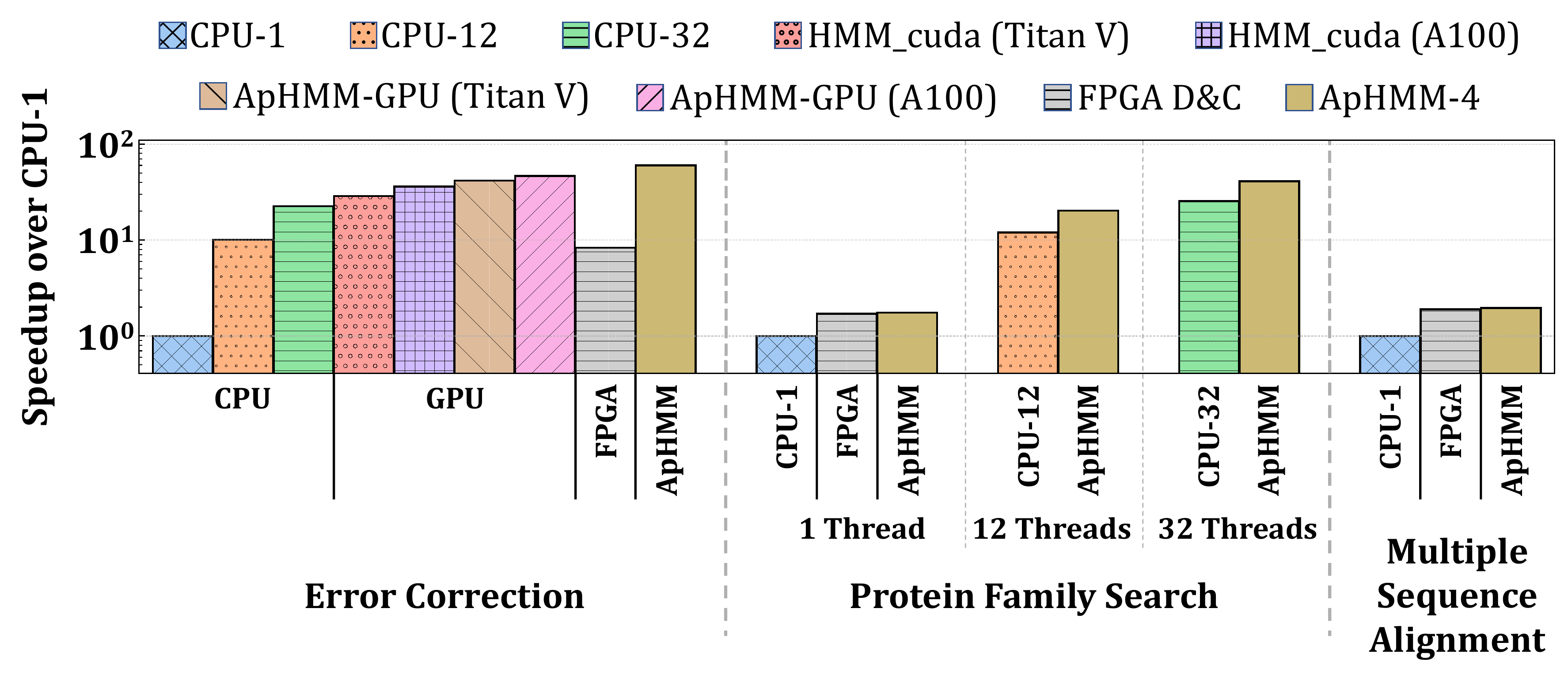}
\caption{Speedups over the single-threaded CPU implementations. In the protein family search, we compare \proposal with each CPU thread separately.}
\label{fig:aphmm_applications}
\end{figure}

Figures~\ref{fig:aphmm_applications} and \ref{fig:baum_energy} show the end-to-end execution time and energy reduction results for error correction, respectively. We make four key observations. 
First, we observe that \proposal is \perrcpuc - \perrcpua, \perrgpud - \perrgpua, and \perrfpga faster than the CPU, GPU, and FPGA implementations of Apollo, respectively.
Second, \proposal reduces the energy consumption by \eerrcpua and \eerrgpud - \eerrgpua compared to the single-threaded CPU and GPU implementations. These two observations are in line with the observations we make in Section~\ref{subsec:baumwelch} as well as the motivation results we describe in Section~\ref{sec:motivation}: Apollo is mainly bounded by the Baum-Welch algorithm, and \proposal accelerates the Baum-Welch algorithm significantly, providing significant performance improvements and energy reductions for error correction. 
We conclude that \proposal significantly improves the energy efficiency and performance of the error correction mainly because the Baum-Welch algorithm constitutes a large portion of the entire use case.

\subsection{Use Case 2: Protein Family Search} \label{subsec:evproteinfamily}

Our goal is to evaluate the performance and energy consumption of \proposal for the protein family search use case, as shown in Figures~\ref{fig:aphmm_applications} and \ref{fig:baum_energy}, respectively. We make three key observations.
First, we observe that \proposal provides speedup by \pprocpuc - \pprocpua, and \pprofpga compared to the CPU and FPGA implementations.
Second, \proposal is \eprocpua more energy efficient than the single-threaded CPU implementation. The speedup ratio that \proposal provides is lower in the protein family search than error correction because 1) \proposal accelerates a smaller portion of the protein family search ($45.76\%$) than error correction ($98.57\%$), and 2) the protein alphabet size (20) is much larger than the DNA alphabet size (4), which increases the DRAM access overhead of \proposal by $12.5\%$. Due to the smaller portion that \proposal accelerates and increased memory accesses, it is expected that \proposal provides lower performance improvements and energy reductions compared to the error correction use case. Third, \proposal can provide better speedup compared to the multi-threaded CPU as a large portion of the parts that \proposal does not accelerate can still be executed in parallel using the same amount of threads, as shown in Figure~\ref{fig:aphmm_applications}.
We conclude that \proposal improves the performance and energy efficiency for the protein family search, while there is a smaller room for acceleration compared to the error correction.

\subsection{Use Case 3: Multiple Sequence Alignment} \label{subsec:evmsa}

Our goal is to evaluate the \proposal's end-to-end performance and energy consumption for the multiple sequence alignment (MSA), as shown in Figures~\ref{fig:aphmm_applications} and \ref{fig:baum_energy}, respectively. We make three key observations.
First, we observe that \proposal performs \pmsacpua and \pmsafpga better than the CPU and FPGA implementations, while \proposal is \emsacpua more energy efficient than the CPU implementation of MSA. We note that the hmmalign tool does not provide the multi-threaded CPU implementation for MSA.
\proposal provides better speedup for MSA than the protein family search because MSA performs more forward and backward calculations ($51.44\%$) than the protein search use case ($45.76\%$), as shown in Figure~\ref{fig:motivation}. Third, \proposal provides slightly better performance than the existing FPGA accelerator (FPGA D\&C) in all applications, even though we ignore the data movement overhead of FPGA D\&C, which suggests that \proposal may perform much better than FPGA D\&C in real systems.
We conclude that \proposal improves the performance and energy efficiency of the MSA use case better than the protein family search.

%% file: tables/breakdown.tex
\begin{tabular}{@{}lr@{}}\toprule
\textbf{Optimization}      & \multicolumn{1}{c}{Speedup ($\times$)} \\\midrule
Histogram Filter                 & 1.07 \\
LUTs                             & 2.48 \\ 
Broadcasting and Partial Compute & 3.39 \\ 
Memoization                      & 1.69 \\\midrule 
Overall                          & 15.20 \\\bottomrule
\end{tabular}

%% file: sections/6_related-work.tex
\section{Related Work}\label{sec:relatedWork}
To our knowledge, this is the first work that provides a flexible and hardware-software co-designed acceleration framework to efficiently and effectively execute the complete Baum-Welch algorithm for pHMMs. In this section, we explain previous attempts to accelerate \emph{HMMs}. Previous works~\cite{ibrahim_reconfigurable_2016, soiman_parallel_2014, huang_hardware_2017, yu_gpu-accelerated_2014, pietras_fpga_2017, eddy_accelerated_2011, ren_fpga_2015, li_improved_2021, wertenbroek_acceleration_2019, banerjee_accelerating_2017, wu_high-throughput_2021, wu_173_2020, jiang_cudampf_2018, quirem_cuda_2011, derrien_hardware_2008, oliver_high_2007, oliver_integrating_2008} mainly focus on specific algorithms and designs of HMMs to accelerate the HMM-based applications. Several works~\cite{ibrahim_reconfigurable_2016, jiang_cudampf_2018, quirem_cuda_2011, derrien_hardware_2008, oliver_high_2007, oliver_integrating_2008} propose FPGA- or GPU-based accelerators for pHMMs to accelerate a different algorithm used in the inference step for pHMMs.
A group of previous works~\cite{huang_hardware_2017, soiman_parallel_2014, ren_fpga_2015, li_improved_2021} accelerates the Forward calculation based on the HMM designs different than pHMMs for FPGAs and supercomputers. HMM\_cuda~\cite{yu_gpu-accelerated_2014} uses GPUs to accelerate the Baum-Welch algorithm for any HMM design. \proposal differs from all of these works as it accelerates the entire Baum-Welch algorithm on pHMMs for more optimized performance, while these works are oblivious to the pHMM design when accelerating the Baum-Welch algorithm.

A related design choice to pHMMs is Pair HMMs. Pair HMMs are useful for identifying differences between DNA and protein sequences. To identify differences, Pair HMMs use states to represent a certain scoring function (e.g., affine gap penalty) or variation type (i.e., insertion, deletion, mismatch, or match) by typically using only one state for each score or difference.
This makes Pair HMMs a good candidate for generalizing pairwise sequence comparisons as they can compare pairs of sequences while being oblivious to any sequence. Unlike pHMMs, Pair HMMs are not built to represent sequences. Thus, Pair HMMs cannot 1)~compare a sequence to a group of sequences and 2)~perform error correction. Pair HMMs mainly target variant calling and sequence alignment problems in bioinformatics. There is a large body of work that accelerates Pair HMMs~\cite{huang_hardware_2017, li_improved_2021, ren_fpga_2015, wertenbroek_acceleration_2019, banerjee_accelerating_2017, wu_high-throughput_2021, wu_173_2020}. \proposal differs from these works as its hardware-software co-design is optimized for pHMMs.

%% file: sections/7_conclusion.tex
\section{Conclusion} \label{sec:conclusion}
We propose \proposal, the first hardware-software co-design framework that accelerates the execution of the entire Baum-Welch algorithm for pHMMs.
\proposal particularly accelerates the Baum-Welch algorithm as it causes a significant computational overhead for important bioinformatics applications. \proposal proposes several hardware-software optimizations to efficiently and effectively execute the Baum-Welch algorithm for pHMMs. The hardware-software co-design of \proposal provides significant performance improvements and energy reductions compared to CPU, GPU, and FPGAs, as \proposal minimizes redundant computations and data movement overhead for executing the Baum-Welch algorithm.
We hope that \proposal enables further future work by accelerating the remaining steps used with pHMMs (e.g., Viterbi decoding) based on the optimizations we provide in \proposal.

%% file: sections/supp.tex
\onecolumn
\setcounter{secnumdepth}{3}
\begin{center}
\textbf{\LARGE Supplementary Material for\\ \ltitle}
\end{center}
\setcounter{section}{0}
\setcounter{equation}{0}
\setcounter{figure}{0}
\setcounter{table}{0}
\setcounter{page}{1}
\makeatletter
\renewcommand{\theequation}{S\arabic{equation}}
\renewcommand{\thetable}{S\arabic{table}}
\renewcommand{\thefigure}{S\arabic{figure}}
\renewcommand{\thesection}{S\arabic{section}}

\newcommand{\TextUnderscore}{\rule{.4em}{.4pt}}

\section{Profile Hidden Markov Models (pHMMs)}\label{suppsec:phmms}
\subsection{High-level Overview}\label{suppsubsec:phmmoverview}
We explain the design of profile Hidden Markov Models (pHMMs). Figure~\ref{fig:phmm} shows the \emph{traditional} structure of pHMMs. To represent a biological sequence in pHMMs and account for differences between the represented sequences and other sequences, pHMMs have a certain graph structure. Visiting nodes, called \emph{states}, via directed edges, called \emph{transitions}, are associated with probabilities to identify differences. To assign a probability for any modification at any sequence position, states are created for each character of the represented sequence. When visited, states \emph{emit} one of the characters from the defined alphabet of the biological sequence (e.g., A, C, T, and G in DNA sequences) with a certain probability. Transitions preserve the correct order of the represented sequences and allow making modifications to thee sequences.

To represent and compare biological sequences, pHMMs are used in three steps.
First, to represent a sequence, pHMM builds the states and transitions by iterating over each character of the sequence. Multiple sequences can also be represented with a single pHMM graph.
A typical pHMM graph includes insertion, match/mismatch, and deletion states for each character of the represented sequence. Connections between states have predefined patterns, as illustrated in Figure~\ref{fig:phmm}. Match states have connections to only match and deletion states of the next character and insertion state of the same character. Deletion states connect to match and deletion states of the next character. Insertion states connect to themselves with a loop and the match state of the next character. The flow from previous to next characters ensures the correct order of the represented sequence in a pHMM graph.

Second, the training step maximizes the similarity score of sequences that are similar to the sequence that the pHMM graph represents. To this end, the training step uses additional input sequences as observation to modify the probabilities of the pHMM. The Baum-Welch algorithm~\citesupp{supp_baum_inequality_1972} is a highly accurate training algorithm for pHMMs.

Third, the inference step aims to either 1)~calculate the similarity score of an input sequence to the sequence represented by a pHMM or 2)~identify the consensus sequence that generates the best similarity score from a pHMM graph. 1)~Calculating the similarity score is useful for applications such as protein family search and MSA. This is because pHMM graphs can avoid making redundant comparisons between sequences by comparing a sequence to a single pHMM graph that represents multiple sequences. Parts of the Baum-Welch algorithm (i.e., the Forward and Backward calculations) can be used in this step for calculating the scores~\citesupp{supp_eddy_accelerated_2011}. 2)~The goal of generating the consensus sequence is to identify the modifications that need to be applied to the represented sequence. These modifications enable error correction tools to identify and correct the errors in DNA sequences. Decoding algorithms such as the Viterbi decoding~\citesupp{supp_viterbi_error_1967} are commonly used for inference from pHMMs~\citesupp{supp_kern_predicting_2013, supp_friedrich_modelling_2006}.

\subsection{Components of pHMMs} \label{suppsubsec:phmmcomponents}
We formally define the pHMM graph structure and its components. We assume that pHMM is a graph, $G(V, A)$, the sequence that the pHMM represents is $S_{G}$, and the length of the sequence is $n_{S_{G}}$. To accurately represent a sequence, pHMMs use four components: 1)~states, 2)~transitions, 3)~emission, and 4)~transition probabilities. We represent the \emph{states} and \emph{transitions} as the members of the sets $V$ and $A$, respectively. First, for each character of sequence $S_{G}$ at position $t$, $S_{G}[t] \in S_{G}$, pHMMs include $3$ consecutive states, $v_{3t}$, $v_{3t+1}$, and $v_{3t+2} \in V$: 1)~match, 2)~insertion, and 3)~deletion states. Each of these states modifies the character $S_{G}[t]$, inserts additional characters after $S_{G}[t]$, or deletes $S_{G}[t]$.
Second, pHMM graphs include transitions from state $v_i$ to state $v_j$, $\alpha_{ij} \in A$, such that the condition $i \leq j$ always holds true to preserve the correct order of characters in $S_{G}$.
Third, to define how probable to observe a certain character when a state is visited, emission probabilities are assigned for each character in a state. These emission probabilities can account for matches and substitutions in match states when comparing a sequence to a pHMM graph. We represent the emission probability of character $c$ in state $v_{i}$ as $e_c(v_{i})$. 
Fourth, to identify the series of states to visit, probabilities are assigned to transitions. We represent the transition probability of a character between states $v_{i}$ and $v_{j}$ as $\alpha_{ij}$. These four main components build up the entire pHMM graph to represent a sequence and calculate the similarity scores when compared to other sequences.


\subsection{Identifying the Modifications} \label{suppsubsec:modifications}

Figure~\ref{fig:phmm} shows three types of modifications that pHMMs can identify, 1)~insertions, 2)~deletions, and 3)~substitutions when comparing the sequence a pHMM represents (i.e., \texttt{PHMM Sequence} in Figure~\ref{fig:phmm}) to other sequences. First, insertion states can identify the characters that are missing from the pHMM sequence at a certain position. For example, Sequence \#1 in Figure~\ref{fig:phmm} includes three additional \texttt{G} characters after \texttt{A}. To identify such insertions, the highlighted insertion state \texttt{I} can be taken three times after visiting the state with label \texttt{A}. Second, deletion states can identify the characters that are deleted from the sequences we compare with the pHMM sequence. Sequence \#2 in Figure~\ref{fig:phmm} provides significant similarity to the pHMM sequence only with a single character missing. To identify the missing character, the highlighted deletion state is visited as it corresponds to deleting the second character in the pHMM sequence, \texttt{C}. Third, match states can identify the characters in sequences different than the character at the same position of a pHMM sequence, which we call substitutions. The states in Figure~\ref{fig:phmm} with DNA letters are match states and show the characters they represent in the corresponding pHMM sequence. The last character of Sequence \#3 is different than the last character of the pHMM sequence in Figure~\ref{fig:phmm}. Such a substitution is identified by visiting the highlighted match state of the last character of the pHMM sequence.


\section{Data distribution and Memory Layout}\label{subsec:datadistribution}
\begin{figure}[ht]
  \centering
  \includegraphics[width=0.7\linewidth]{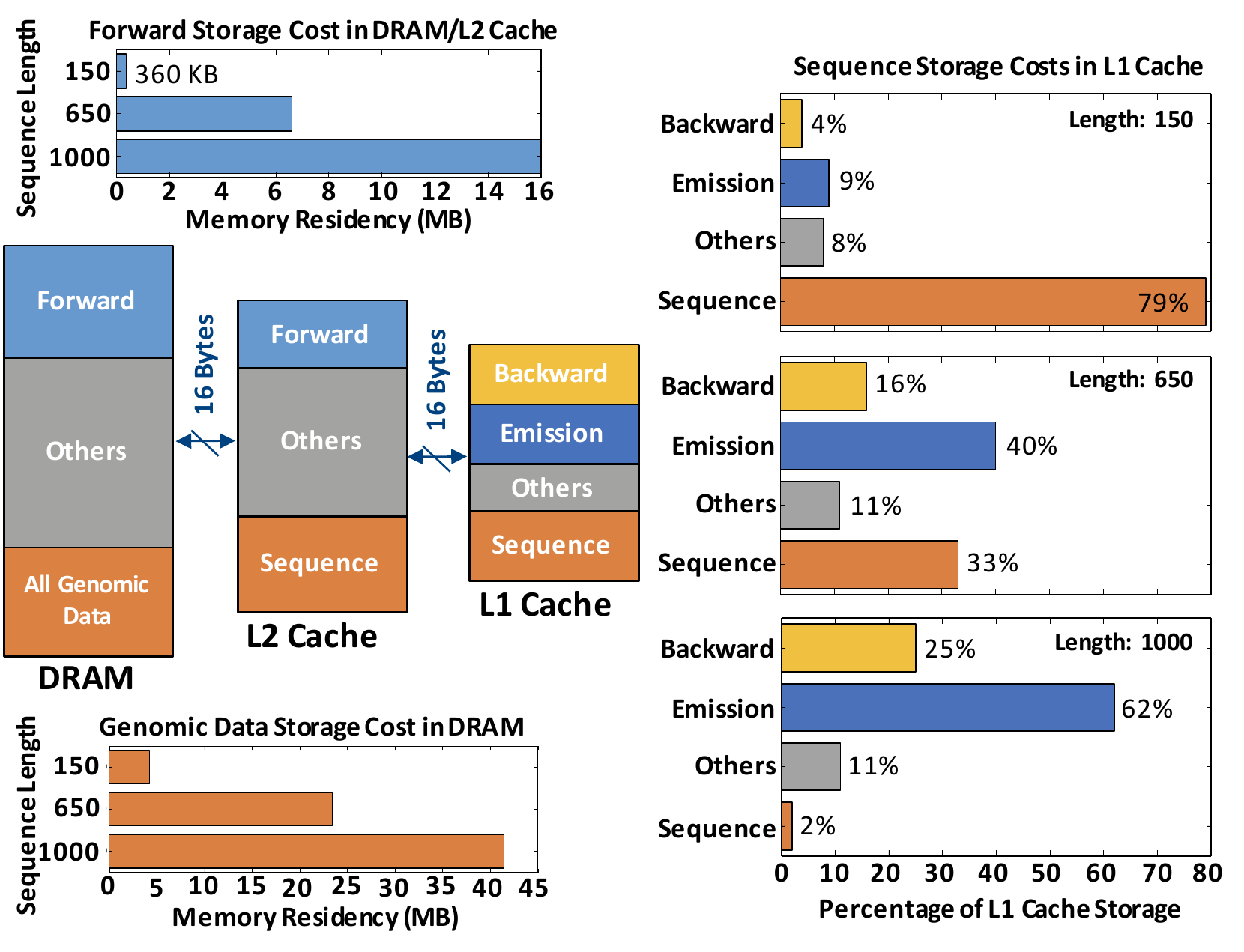}
    \caption{Data distribution across memory hierarchy.}
    \label{fig:l1_data_distribution}
\end{figure}

To efficiently implement genomic sequence execution in a memory-constrained environment, \proposal distributes several types of data utilizing multiple levels of memory hierarchy: DRAM, SRAM-based 4-banked L2 and L1 cache, and on-chip scratchpad and registers. The entire genomic data set is traditionally large and is typically stored in DRAM, with smaller subsets of the data fetched into the L2 and L1 cache as needed. The L1 and L2 caches are divided into multiple sections using an SRAM-based 4-banked cache, with each bank dedicated to a specific type of data as shown in Figure~\ref{fig:l1_data_distribution}. The division of memory into these blocks is not hard-coded, and each section can be dynamically resized as needed. \proposal uses an additional 2 bits to label these four sections in memory blocks.

DRAM and caches mainly hold 1)~chunked sequences that can be directly processed by the \proposal Core, 2)~Forward and Backward values, 3)~emission probabilities, and 4)~other temporary results generated by the \proposal Core. 
First, to store large sequences using memory-constrained resources and enable better parallelism, the sequences are divided into \emph{chunks} of sequence lengths ranging from 150 to 1,000 characters. This is designed to represent both sequencing reads and almost all protein sequences, as these protein sequences are mostly smaller than 1,000 characters~\citesupp{supp_brocchieri_protein_2005}. For longer sequences, a sequence may be chunked into small pieces while preserving the relative order between sequences. A previous analysis shows that chunking does not degrade the accuracy of the training and inference steps~\citesupp{supp_firtina_apollo_2020}. \proposal uses L1 cache of 128KB to support a larger spectrum of sequence lengths ranging between 150-1000 characters. Figure~\ref{fig:l1_data_distribution} shows the size of different Baum-Welch parameters that must be stored in memory based on the sequence length and the details for efficiently storing the data across the memory hierarchy. 
Second, \proposal stores Forward and Backward values across different levels of memory hierarchies. \proposal stores the entire Forward values in DRAM and fetches them into L2 cache as required. Since Backward values are broadcasted without fully computing them, these values are stored in the L1 cache to be broadcasted in the next timestamp.
Third, unlike the transition probabilities stored in scratchpad in a \proposal-core, \proposal stores emission probabilities in L1 cache as numerators and denominators can be calculated independently, providing opportunities for parallel computation while requiring larger memory space for larger alphabet sizes (Section~\ref{subsec:computeblock}) and sequence lengths (Figure~\ref{fig:l1_data_distribution}). Fourth, all the other temporary results are mainly parameters inputted to the \proposal-core at each time step. These values are usually the information regarding the next execution step per state (e.g., per base).

Our key observation from the space requirements of various chunk sizes is that the size of Baum-Welch parameters grows as the sequence length increases. Thus, increased chunk length reduces the number of sequences that L1 cache can hold. This does not cause frequent data load of sequences from DRAM or the L2 cache as longer sequences occupy the \proposal Core usually for a longer duration, compensating for the fewer read sequences stored in L1.

\section{System Mapping and Execution Flow}\label{subsec:sysmapping}
\begin{figure}[ht]
  \centering
  \includegraphics[width=0.5\linewidth]{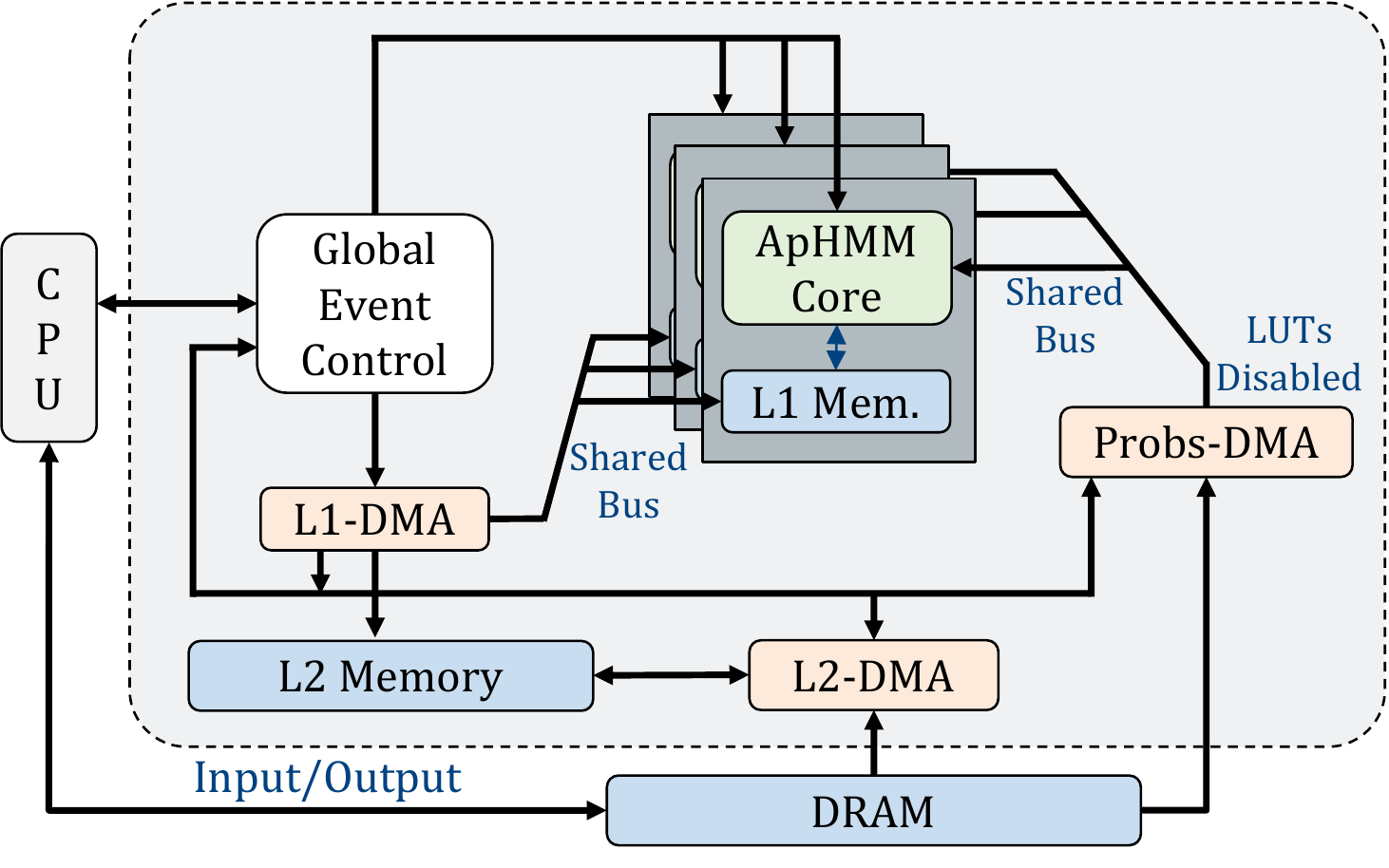}
    \caption{System integration of the \proposal Core.}
    \label{fig:system_mapping}
\end{figure}

We show a system-level scale-up version of the \proposal Core in Figure~\ref{fig:system_mapping}. \proposal uses the L2-DMA table to load the data into the L2 cache and the L1-DMA table to write the corresponding data into the L1 cache per \proposal Core according to the data distribution, as described in Section~\ref{subsec:datadistribution}. \proposal enables Probs-DMA to load the transition probabilities from DRAM to the local memory when the \luts are not utilized, as discussed in Section~\ref{subsec:computeblock}. In such a scenario, local memory inside the \pe is loaded with appropriate transition probability data to perform the multiplications without using \luts.

\begin{figure}[ht]
  \centering
  \includegraphics[width=0.85\linewidth]{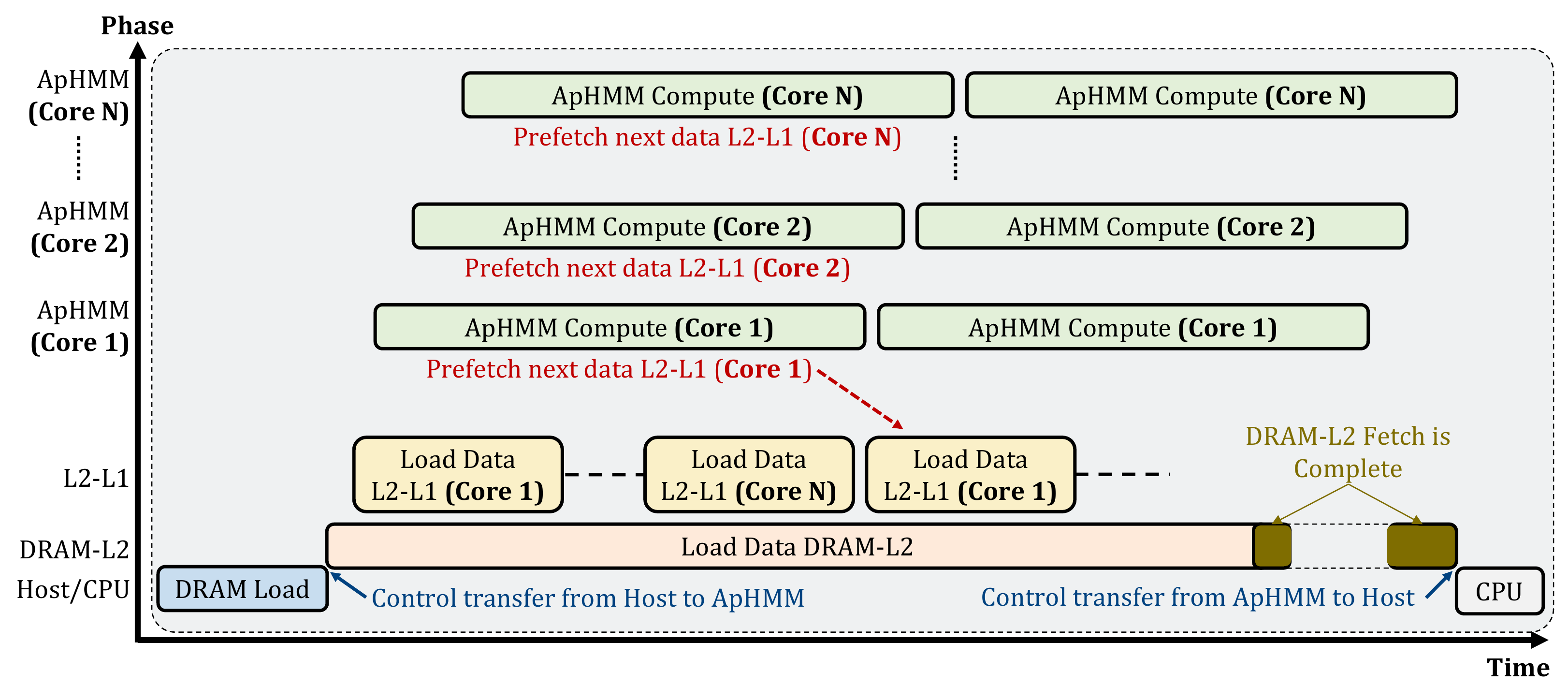}
    \caption{Control and execution flow of \proposal Cores.}
    \label{fig:exec_flow}
\end{figure}

We present the execution flow of the system with multi-\proposal Core in Figure~\ref{fig:exec_flow}. The operation starts with the host loading the data into DRAM and issuing DMA across various memory hierarchies through a global event control. Each \proposal Core can start asynchronously, and near the completion of all reads from L1, hardware sets a flag for fetching the next set of sequences from L2. Similarly, counter-based signaling tells L2 to fetch the next set of sequences from DRAM. Once all reads are issued, \proposal sends a completion signal and releases the control back to the host.

\clearpage

\let\noopsort\undefined
\let\printfirst\undefined
\let\singleletter\undefined
\let\switchargs\undefined

\bibliographystylesupp{IEEEtran}
\bibliographysupp{main}